\documentclass [10pt,twocolumn,oneside,a4paper] {article}
\usepackage{epsf}
\begin{document}
\onecolumn
\title {\textbf{Weak Interaction Rates Of $\mathbf{sd}$- Shell Nuclei In Stellar Environment Calculated in the Proton-Neutron Quasiparticle Random Phase Approximation} }
\author{\textbf{Jameel-Un-Nabi}  \and
\textbf {Hans-Volker Klapdor-Kleingrothaus} \\Max-Planck-Institut f\"ur Kernphysik, 69029 Heidelberg, Germany}
\normalsize
\maketitle
\begin{abstract}
Allowed weak interaction rates for $\mathit{sd}$- shell nuclei in stellar environment are calculated using a generalized form of proton-neutron quasiparticle RPA model with separable Gamow-Teller forces. Twelve different weak rates are calculated for each nucleus as a function of temperature and density. The stellar weak rates are calculated over a wide range of densities (10$\leq$ $\rho Y_{e}$ (gcm$^{-3}$) $ \leq$ 10$^{11}$) and temperatures (10$^{7}$ $\leq$ T(K) $\leq$ 30 $\times$ 10$^{9}$). This project consists of calculation of weak rates for a total of 709 nuclei with masses ranging from A = 18 to 100. This paper contains calculated weak rates for sd-shell nuclei. The calculated capture and decay rates take into consideration the latest experimental energy levels and $ft$ value compilations. The results are also compared with earlier works. Particle emission processes from excited states, previously ignored, are taken into account, and are found to significantly affect some $\beta$ decay rates.
\end{abstract}
\clearpage
\twocolumn
\section {\normalsize INTRODUCTION}
There are many interesting problems in astrophysics which require $e^{\pm}$ -capture and $\beta^{\pm}$ -decay rates as input parameters. These rates are used in numerical simulations for nuclear astrophysical problems. In the course of development of a star, the weak interaction has several crucial effects. They initiate the gravitational collapse of the core of a massive star triggering a supernova explosion, play a key role in neutronisation of the core material, and, affect the formation of heavy elements above iron via the r-process at the final stage of the supernova explosion. The weak interaction also largely determines the mass of the core, and thus the strength and fate of the shock wave formed by the supernova explosion.

Precise knowledge of the $\beta$ decay of neutron-rich nuclei is crucial to an understanding of the r-process. Both the element distribution on the r-path, and the resulting final distribution of stable elements are highly sensitive to the $\beta$ decay properties of the neutron-rich nuclei involved in the process. This was first pointed out by Klapdor [1,2,3] (see also [4]).There are about 6000 nuclei between the $\beta$ stability line and the neutron drip line. Most of these nuclei cannot be produced in terrestrial laboratories and one has to rely on theoretical extrapolations in respect of beta decay properties. Calculations of beta decay rates for all nuclei far from stability by microscopic nuclear theory were first performed by Klapdor et al.[5], and then complemented and refined by Staudt et al. [6,7] and Hirsch et al. [8]. Recent studies by Homma et al. [9] have shown that the best extrapolations to nuclei far from stability to date still are given by [6]. These weak interaction rates led to a better understanding of the r-process. However there was a need to go to domains of high temperature and density scales where the weak interaction rates are of decisive importance in studies of the stellar evolution.

Processes where $\beta$ transitions occur at high temperatures and densities (T$ \sim$ 10$^{9}$ K, $\rho Y_{e}$ $\geq$ 10$^{3}$ gcm$^{-3}$) include, besides those mentioned above -- the initiation of the collapse of the O + Ne + Mg core of 8-10 M$_{\odot}$ stars [10], the growth of the mass of the iron core of M $\geq$ 10 M$_{\odot}$ stars -- the element synthesis problem, like the synthesis of iron-group elements in carbon-detonation supernova models [11], the hot CNO-cycle [12,13], where at temperatures beyond 10$^{8}$ K and densities beyond 1 gcm$^{-3}$, $\beta^{+}$ decays are much slower than thermonuclear reactions. Some other examples include the s-process [14,15], and the p-process [16,17].

The first extensive effort to tabulate the weak interaction rates at high temperatures and densities, where decays from excited states of the parent nuclei become relevant, was done by Fuller, Fowler, and Newmann (FFN) [18]. We will refer to such rates as stellar rates throughout this paper. FFN calculated the stellar weak interaction rates over a wide range of densities and temperatures (10$\leq$ $\mathit{\rho Y_{e}}$ (gcm$^{-3}$) $ \leq$ 10$^{11}$, 10$^{7}$ $\leq$ T(K) $\leq$ 10$^{11}$). The Gamow-Teller (GT) strength and excitation energies were calculated using a zero-order shell model. They also incorporated the experimental data available at that time. Matrix elements of Brown et al. [19] were used for unmeasured GT strengths. When these were also not available FFN assumed an average log $\mathit{ft}$ value of 5.0 for their calculation of weak rates. They calculated the weak rates for 226 nuclei.

Beta transition rates in stellar matter were also calculated by Takahashi et al. [20] using the so-called gross theory of beta decay. The shortfall in this theory is that the nuclear structure details of the individual nuclei are not taken into account and only a statistical description of the $\beta$ strength function is assumed. This then only reproduces some global properties of the nuclei.

Oda et al. (OHMTS) [21] did an extensive calculation of stellar weak interaction rates of $\mathit{sd}$- shell nuclei in full ($\mathit{sd}$)$^{n}$- shell model space. They used the effective interaction of Wildenthal [19,22,23] and the available experimental compilations for their calculations. They also compared their calculated rates with those of FFN and in certain cases they reported differences in the rates up to two order of magnitude and more. 

Proton-neutron quasiparticle random-phase-approximation (pn-QRPA) has been shown to be a good microscopic theory for the  calculation of beta decay half-lives. Bender et al. [24] and Staudt et al [6,7] used the QRPA to calculate the $\beta$$^{-}$ decay half-lives of nuclei far from stability and obtained good agreement with experimental decay rates. Muto et al. [25] then extended this model to treat transitions from nuclear excited states. Keeping in view the success of pn-QRPA theory in calculating terrestrial decay rates, we used this extended model to calculate for the first time the weak interaction rates in stellar matter using pn-QRPA theory. Particle emission processes from excited states, which were not considered by OHMTS and FFN, are taken into account. Our calculation of stellar rates for $\mathit{sd}$- shell nuclei shows significant differences, especially for decay rates, compared to the earlier work of OHMTS and FFN. This comparison is treated in detail in subsection 5.2. The use of pn-QRPA theory at finite temperatures also enables us to calculate the weak interaction rates of $\mathit{fp}$- shell nuclei. To date there is no calculation of weak rates for $\mathit{fp}$- shell nuclei in hot and dense matter taking into account details of the nuclear structure. The work here is the beginning of our extensive calculations of stellar rates. We start with the $\mathit{sd}$- shell nuclei. Our calculation is more extensive than the previous calculations of OHMTS, both in number of nuclei considered and number of weak processes calculated. We calculated the weak interaction rates at precisely the same temperature and density grid points as OHMTS in order to present a direct comparison of the two results where possible.

In section 2, we present the formalism of rate calculations for various nuclear weak processes. The description of pn-QRPA and its extended model to calculate the nuclear matrix elements is dealt with in section 3. Latest experimental data is also incorporated in our work and section 4 describes the procedure involved. Results and discussions are treated in section 5. Here we also compare our work with the earlier work of OHMTS and FFN. Section 6 then summarizes our results.

The application of this formalism to the calculation of $\mathit{fp}$- shell nuclei is under way and will be published shortly in a later paper.
\section{\normalsize FORMALISM}
\subsection{Assumptions}
The following main assumptions are made:

(1) Only allowed GT and superallowed Fermi transitions are calculated since contributions from forbidden transitions are relatively negligible.

(2) We assume that the temperature is high enough to ionize the atoms completely. The electrons are not bound anymore to the nucleus and obey the Fermi-Dirac distribution. Neutrino and anti-neutrino capture is not taken into account.

(3) Particle emissions from excited states are not ignored and the effects of separation energy of protons ($S_{p}$) and neutrons ($S_{n}$) on the stellar rates are considered.

(4) All excited states, with energy less than $S_{p}$ (or $S_{n}$), decay directly to the ground state through $\gamma$ transitions. The $\beta$ decay of possible isomeric states is not taken into account.
\subsection{Beta Decay in Stellar Matter}
The weak decay rate from the $\mathit{i}$th state of the parent to the $\mathit{j}$th state of the daughter nucleus is given by \footnote{Throughout section 2 we use natural units $(\hbar=c=m_{e}=1)$, unless otherwise stated, where $m_{e}$ is the electron mass.}
\begin{equation}
\lambda_{ij} =ln2 \frac{f_{ij}(T,\rho,E_{f})}{(ft)_{ij}}
\end{equation}
where $(ft)_{ij}$ is related to the reduced transition probability $B_{ij}$ of the nuclear transition by
\begin{equation}
(ft)_{ij}=D/B_{ij}
\end{equation}
The $D$ appearing in Eq. ~(2) is a compound expression of physical constants,
\begin{equation}
D=\frac{2ln2\hbar^{7}\pi^{3}}{g_{V}^{2}m_{e}^{5}c^{4}}
\end{equation}
and,
\begin{equation}
B_{ij}=B(F)_{ij}+(g_{A}/g_{V})^2 B(GT)_{ij}
\end{equation}
where B(F) and B(GT) are reduced transition probabilities of the Fermi and ~Gamow-Teller (GT) transitions respectively,
\begin{equation}
B(F)_{ij} = \frac{1}{2J_{i}+1} \mid<j \parallel \sum_{k}t_{\pm}^{k} \parallel i> \mid ^{2}
\end{equation} 
\begin{equation}
B(GT)_{ij} = \frac{1}{2J_{i}+1} \mid <j \parallel \sum_{k}t_{\pm}^{k}\vec{\sigma}^{k} \parallel i> \mid ^{2}
\end{equation}

In Eq. ~(6), $\vec{\sigma}^{k}$ is the spin operator and $t_{\pm}^{k}$ stands for the isospin raising and lowering operator. We take the value of D=6295 s and the ratio of the axial-vector $(g_{A})$ to the vector $(g_{V})$ coupling constant as 1.254. The calculation of nuclear matrix elements is dealt with in section 3.

The phase space integral $(f_{ij})$ is an integral over total energy,
\begin{equation}
f_{ij} = \int_{1}^{w_{m}} w \sqrt{(w^{2}-1)} (w_{m}-w)^{2} F(\pm Z,w) (1-G_{\mp}) dw
\end{equation}
for electron (\textit{upper signs}) or positron (\textit{lower signs}) emission, or by
\begin{equation}
f_{ij} = \int_{w_{l}}^{\infty} w \sqrt{(w^{2}-1)} (w_{m}+w)^{2} F(\pm Z,w) G_{\mp} dw 
\end{equation}
for continuum positron (\textit{lower signs}) or electron (\textit{upper signs}) capture.

In Eqs. (7) and (8), $w$ is the total kinetic energy of the electron including its rest mass, $w_{l}$ is the total capture threshold energy (rest+kinetic) for positron (or electron) capture. One should note that if the corresponding electron (or positron) emission total energy, $w_{m}$, is greater than -1, then \\$w_{l}=1$, and if it is less than or equal to 1, then $w_{l}=\mid w_{m} \mid$. $w_{m}$ is the total $\beta$ -decay energy,
\begin{equation}
w_{m} = m_{p}-m_{d}+E_{i}-E_{j}
\end{equation}
where $m_{p}$ and $E_{i}$ are mass and excitation energies of the parent nucleus, and $m_{d}$ and $E_{j}$ of the daughter nucleus, respectively.

$G_{+}$ and $G_{-}$ are the positron and electron distribution functions, respectively. Assuming that the electrons are not in a bound state, these are the Fermi-Dirac distribution functions,
\begin{equation}
G_{-} = [exp (\frac{E-E_{f}}{kT})+1]^{-1}
\end{equation}
\begin{equation}
G_{+} = [exp (\frac{E+2+E_{f}}{kT})+1]^{-1}
\end{equation}
Here $E=(w-1)$ is the kinetic energy of the electrons, $E_{f}$ is the Fermi energy of the electrons, $T$ is the temperature, and $k$ is the Boltzmann constant.

In our calculations, the inhibition of the final neutrino phase space is never appreciable enough that neutrino (or anti-neutrino) distribution functions had to be taken into consideration. $F(\pm Z,w)$ are the Fermi functions and are calculated according to the procedure adopted by Gove and Martin [26].

The number density of electrons associated with protons and nuclei is $\rho Y_{e} N_{A}$, where $\rho$ is the baryon density, $Y_{e}$ is the ratio of electron number to the baryon number, and $N_{A}$ is the Avogadro's number.
\begin{equation}
\rho Y_{e} = \frac{1}{\pi^{2}N_{A}}(\frac {m_{e}c}{\hbar})^{3} \int_{0}^{\infty} (G_{-}-G_{+}) p^{2}dp 
\end{equation}
where $p=(w^{2}-1)^{1/2}$ is the electron or positron momentum, and Eq. ~(12) has the units of \textit{moles $cm^{-3}$}. This equation is used for an iterative calculation of Fermi energies for selected values of $\rho Y_{e}$ and $T$.

There is a finite probability of occupation of parent excited states in the stellar environment as a result of the high temperature in the interior of massive stars. Weak interaction rates then also have a finite contribution from these excited states. The occupation probability of a state $i$ is calculated on the assumption of thermal equilibrium,
\begin{equation}
P_{i} = \frac {(2J_{i}+1)exp(-E_{i}/kT)}{\sum_{i=1}(2J_{i}+1)exp(-E_{i}/kT)}
\end{equation}
where $J_{i}$ and $E_{i}$ are the angular momentum and excitation energy of the state $i$, respectively.

Unfortunately we cannot calculate the $J_{i}$ in QRPA theory and hence we are forced to modify Eq. ~(13) as
\begin{equation}
P_{i} = \frac {exp(-E_{i}/kT)}{\sum_{i=1}exp(-E_{i}/kT)}
\end{equation}
This approximation is a compromise and can be justified when we take into consideration the uncertainty in the calculation of $E_{i}$ which easily over-sheds the uncertainty in calculating the values of $J_{i}$ in the above Eq. ~(13).

The rate per unit time per nucleus for any weak process is then given by
\begin{equation}
\lambda = \sum_{ij}P_{i} \lambda_{ij}
\end{equation}
We carry out this summation over all initial and final states until satisfactory convergence in our rate calculations is achieved.

The neutrino energy loss rates are calculated using the same formalism except that the phase space integral is replaced by
\begin{equation}
f_{ij}^{\nu} = \int_{1}^{w_{m}} w \sqrt{(w^{2}-1)} (w_{m}-w)^{3} F(\pm Z,w) (1-G_{\mp}) dw
\end{equation}
for electron (\textit{upper signs}) or positron (\textit{lower signs}) emission, or by
\begin{equation}
f_{ij}^{\nu} = \int_{w_{l}}^{\infty} w \sqrt{(w^{2}-1)} (w_{m}+w)^{3} F(\pm Z,w) G_{\mp} dw  
\end{equation}
for continuum positron (\textit{lower signs}) or electron (\textit{upper signs}) capture.

We calculate the proton energy rate from the daughter nucleus, whenever $S_{p}$ $<$ $S_{n}$, by
\begin{equation}
\lambda^{p} = \sum_{ij}P_{i}\lambda_{ij}(E_{j}-S_{p})
\end{equation}
for all $E_{j} > S_{p}$, whereas for all $E_{j} \leq S_{p}$ we calculate the $\gamma$ heating rate,
\begin{equation}
\lambda^{\gamma} = \sum_{ij}P_{i}\lambda_{ij}E_{j}
\end{equation}
If on the other hand, $S_{n} < S_{p}$, then we calculate the neutron energy rate from the daughter nucleus by
\begin{equation}
\lambda^{n} = \sum_{ij}P_{i}\lambda_{ij}(E_{j}-S_{n})
\end{equation}
for all $E_{j} > S_{n}$, and for all $E_{j} \leq S_{n}$ we calculate the $\gamma$ heating rate as in Eq. ~(19).  

The probability of $\beta$-delayed proton (neutron) emission is calculated by
\begin{equation}
P^{p(n)} = \frac{\sum_{ij\prime}P_{i}\lambda_{ij\prime}}{\sum_{ij}P_{i}\lambda_{ij}}
\end{equation}
where $j\prime$ are states in the daughter nucleus for which $E_{j\prime} > S_{p} (S_{n})$. In Eqs. [(18)-(21)] $\lambda_{ij(\prime)}$ is the sum of the electron capture and positron decay rates, or the sum of positron capture and electron decay rates, for the transition $i$ $\rightarrow$ $j(j\prime)$.
\section{\normalsize CALCULATION OF NUCLEAR MATRIX ELEMENTS}
\subsection{Extended Model of pn-QRPA}
Application of the pn-QRPA theory for the calculation of beta decay half-lives is not new. As such we will not be explaining the general form of pn-QRPA. Interested readers are referred to, eg., [25,27,28]. By taking separable Gamow-~Teller forces the energy eigenvalue problem of the RPA equation reduces to finding roots of an algebraic equation of fourth order [25]. Here we only quote the extension of this model to include transitions from parent excited states [25].

The parent excited states can be constructed as phonon-correlated multi-quasi-particle states. The transition amplitudes between the multi-quasi-particle states can be reduced to those of single-quasi-particle states as shown below.

Excited states of an even-even nucleus are two-proton quasi-particle (q.p.) states and two-neutron q.p. states. Transitions from these initial states are possible to final proton-neutron q.p. pair states in the odd-odd daughter nucleus. The transition amplitudes and their reduction to correlated ($c$) one-q.p.states is given by
\begin{eqnarray*}
<p^{f}n_{c}^f \mid t_{\pm}\sigma_{-\mu} \mid p_{1}^{i}p_{2c}^{i}> \hspace{0.55in} \\
 = -\delta (p^{f},p_{2}^{i}) <n_{c}^{f} \mid t_{\pm}\sigma_{-\mu} \mid p_{1c}^{i}> \\
+\delta (p^{f},p_{1}^{i}) <n_{c}^{f} \mid t_{\pm}\sigma_{-\mu} \mid p_{2c}^{i}> \\
(22)
\end{eqnarray*}
\begin{eqnarray*}
<p^{f}n_{c}^f \mid t_{\pm}\sigma_{\mu} \mid n_{1}^{i}n_{2c}^{i}> \hspace{0.55in} \\
 = +\delta (n^{f},n_{2}^{i}) <p_{c}^{f} \mid t_{\pm}\sigma_{\mu} \mid n_{1c}^{i}> \\
-\delta (n^{f},n_{1}^{i}) <p_{c}^{f} \mid t_{\pm}\sigma_{\mu} \mid n_{2c}^{i}> \\
(23)
\end{eqnarray*}
where $\mu$ = -1, 0, 1, are the spherical components of the spin operator.

For an odd-neutron even-proton nucleus, the excited states can be constructed\\
(1) by lifting the odd neutron from ground state to excited states (one-q.p. state),\\
(2) by three-neutron states, corresponding to excitation of a neutron, or,\\
(3) by one-neutron two-proton states, corresponding to excitation of a proton.

Here we present only the formulae for multi-q.p transitions and their reduction to correlated ($c$) one-q.p. states,
\begin{eqnarray*}
<p_{1}^{f}n_{1}^{f}n_{2c}^{f} \mid t_{\pm}\sigma_{\mu} \mid n_{1}^{i}n_{2}^{i}n_{3c}^{i}> \hspace{0.7in} \\
 = \delta (n_{1}^{f},n_{2}^{i}) \delta (n_{2}^{f},n_{3}^{i}) <p_{1c}^{f} \mid t_{\pm}\sigma_{\mu} \mid n_{1c}^{i}> \nonumber \\
 - \delta (n_{1}^{f},n_{1}^{i}) \delta (n_{2}^{f},n_{3}^{i}) <p_{1c}^{f} \mid t_{\pm}\sigma_{\mu} \mid n_{2c}^{i}> \nonumber \\
 + \delta (n_{1}^{f},n_{1}^{i}) \delta (n_{2}^{f},n_{2}^{i}) <p_{1c}^{f} \mid t_{\pm}\sigma_{\mu} \mid n_{3c}^{i}> \\
(24)
\end{eqnarray*}
\begin{eqnarray*}
<p_{1}^{f}n_{1}^{f}n_{2c}^{f} \mid t_{\pm}\sigma_{-\mu} \mid p_{1}^{i}p_{2}^{i}n_{1c}^{i}> \hspace{0.7in}\\
 = \delta (p_{1}^{f},p_{2}^{i})[ \delta (n_{1}^{f},n_{1}^{i}) <n_{2c}^{f} \mid t_{\pm}\sigma_{-\mu} \mid p_{1c}^{i}>  \\
 - \delta (n_{2}^{f},n_{1}^{i}) <n_{1c}^{f} \mid t_{\pm}\sigma_{-\mu} \mid p_{1c}^{i}>] \\ 
 -\delta (p_{1}^{f},p_{1}^{i})[ \delta (n_{1}^{f},n_{1}^{i}) <n_{2c}^{f} \mid t_{\pm}\sigma_{-\mu} \mid p_{2c}^{i}> \\
  - \delta (n_{2}^{f},n_{1}^{i}) <n_{1c}^{f} \mid t_{\pm}\sigma_{-\mu} \mid p_{2c}^{i}>] \\
(25)
\end{eqnarray*}
\begin{eqnarray*}
<p_{1}^{f}p_{2}^{f}p_{3c}^{f} \mid t_{\pm}\sigma_{\mu} \mid p_{1}^{i}p_{2}^{i}n_{1c}^{i}> \hspace{0.7in}\\
 = \delta (p_{2}^{f},p_{1}^{i}) \delta (p_{3}^{f},p_{2}^{i}) <p_{1c}^{f} \mid t_{\pm}\sigma_{\mu} \mid n_{1c}^{i}> \\
 - \delta (p_{1}^{f},p_{1}^{i}) \delta (p_{3}^{f},p_{2}^{i}) <p_{2c}^{f} \mid t_{\pm}\sigma_{\mu} \mid n_{1c}^{i}> \\
 + \delta (p_{1}^{f},p_{1}^{i}) \delta (p_{2}^{f},p_{2}^{i}) <p_{3c}^{f} \mid t_{\pm}\sigma_{\mu} \mid n_{1c}^{i}> \\
(26)
\end{eqnarray*}

Similarly low-lying states of an odd-proton even-neutron nucleus can be constructed,\\
(1) by exciting the odd proton from ground state (one-q.p. states),\\
(2) by three-proton states, corresponding to excitation of a proton, or,\\
(3) by one-proton two-neutron states, corresponding to excitation of a neutron.

The multi-q.p. transitions can again be reduced to correlated ($c$) one-q.p. states,
\begin{eqnarray*}
<p_{1}^{f}p_{2}^{f}n_{1c}^{f} \mid t_{\pm}\sigma_{-\mu} \mid p_{1}^{i}p_{2}^{i}p_{3c}^{i}> \hspace{0.7in} \\
 = \delta (p_{1}^{f},p_{2}^{i}) \delta (p_{2}^{f},p_{3}^{i}) <n_{1c}^{f} \mid t_{\pm}\sigma_{-\mu} \mid p_{1c}^{i}> \\
 - \delta (p_{1}^{f},p_{1}^{i}) \delta (p_{2}^{f},p_{3}^{i}) <n_{1c}^{f} \mid t_{\pm}\sigma_{-\mu} \mid p_{2c}^{i}> \\
 + \delta (p_{1}^{f},p_{1}^{i}) \delta (p_{2}^{f},p_{2}^{i}) <n_{1c}^{f} \mid t_{\pm}\sigma_{-\mu} \mid p_{3c}^{i}> \\
(27)
\end{eqnarray*}
\begin{eqnarray*}
<p_{1}^{f}p_{2}^{f}n_{1c}^{f} \mid t_{\pm}\sigma_{\mu} \mid p_{1}^{i}n_{1}^{i}n_{2c}^{i}> \hspace{0.7in} \\
 = \delta (n_{1}^{f},n_{2}^{i})[ \delta (p_{1}^{f},p_{1}^{i}) <p_{2c}^{f} \mid t_{\pm}\sigma_{\mu} \mid n_{1c}^{i}> \\
 - \delta (p_{2}^{f},p_{1}^{i}) <p_{1c}^{f} \mid t_{\pm}\sigma_{\mu} \mid n_{1c}^{i}>] \nonumber \\ 
-\delta (n_{1}^{f},n_{1}^{i})[ \delta (p_{1}^{f},p_{1}^{i}) <p_{2c}^{f} \mid t_{\pm}\sigma_{\mu} \mid n_{2c}^{i}> \\
 - \delta (p_{2}^{f},p_{1}^{i}) <p_{1c}^{f} \mid t_{\pm}\sigma_{\mu} \mid n_{2c}^{i}>]\\
(28)
\end{eqnarray*}
\begin{eqnarray*}
<n_{1}^{f}n_{2}^{f}n_{3c}^{f} \mid t_{\pm}\sigma_{-\mu} \mid p_{1}^{i}n_{1}^{i}n_{2c}^{i}> \hspace{0.7in} \\
=\delta (n_{2}^{f},n_{1}^{i}) \delta (n_{3}^{f},n_{2}^{i}) <n_{1c}^{f} \mid t_{\pm}\sigma_{-\mu} \mid p_{1c}^{i}> \\
\mbox{} -\delta (n_{1}^{f},n_{1}^{i}) \delta (n_{3}^{f},n_{2}^{i}) <n_{2c}^{f} \mid t_{\pm}\sigma_{-\mu} \mid p_{1c}^{i}> \\
\mbox{} +\delta (n_{1}^{f},n_{1}^{i}) \delta (n_{2}^{f},n_{2}^{i}) <n_{3c}^{f} \mid t_{\pm}\sigma_{-\mu} \mid p_{1c}^{i}> \\
(29) 
\end{eqnarray*}

States in an odd-odd nucleus are expressed in q.p. transformation by two-q.p. states (proton-neutron pair states) or by four-q.p. states (two-proton or two-neutron q.p. states). Reduction of two-q.p. states into correlated ($c$) one-q.p. states are given as
\begin{eqnarray*}
<p_{1}^{f}p_{2c}^{f} \mid t_{\pm}\sigma_{\mu} \mid p^{i}n_{c}^{i}> \hspace{0.4in} \\
= \delta(p_{1}^{f},p^{i}) <p_{2c}^{f} \mid t_{\pm}\sigma_{\mu} \mid n_{c}^{i}> \\
- \delta(p_{2}^{f},p^{i}) <p_{1c}^{f} \mid t_{\pm}\sigma_{\mu} \mid n_{c}^{i}> \\
(30)
\end{eqnarray*}
\begin{eqnarray*}
<n_{1}^{f}n_{2c}^{f} \mid t_{\pm}\sigma_{-\mu} \mid p^{i}n_{c}^{i}> \hspace{0.4in} \\
= \delta(n_{2}^{f},n^{i}) <n_{1c}^{f} \mid t_{\pm}\sigma_{-\mu} \mid p_{c}^{i}> \\
- \delta(n_{1}^{f},n^{i}) <n_{2c}^{f} \mid t_{\pm}\sigma_{-\mu} \mid p_{c}^{i}> \\
(31)
\end{eqnarray*}
while the four-q.p. states are simplified as
\begin{eqnarray*}
<p_{1}^{f}p_{2}^{f}n_{1}^{f}n_{2c}^{f} \mid t_{\pm}\sigma_{-\mu} \mid p_{1}^{i}p_{2}^{i}p_{3}^{i}n_{1c}^{i}> \hspace{0.3in} \\
=\delta (n_{2}^{f},n_{1}^{i})[ \delta (p_{1}^{f},p_{2}^{i})\delta (p_{2}^{f},p_{3}^{i}) \hspace{0.7in} \\
<n_{1c}^{f} \mid t_{\pm}\sigma_{-\mu} \mid p_{1c}^{i}> \hspace{1.15in} \\
-\delta (p_{1}^{f},p_{1}^{i}) \delta (p_{2}^{f},p_{3}^{i}) <n_{1c}^{f} \mid t_{\pm}\sigma_{-\mu} \mid p_{2c}^{i}> \\
+\delta (p_{1}^{f},p_{1}^{i}) \delta (p_{2}^{f},p_{2}^{i}) <n_{1c}^{f} \mid t_{\pm}\sigma_{-\mu} \mid p_{3c}^{i}>] \\
-\delta (n_{1}^{f},n_{1}^{i})[ \delta (p_{1}^{f},p_{2}^{i})\delta (p_{2}^{f},p_{3}^{i}) \hspace{0.8in} \\
<n_{2c}^{f} \mid t_{\pm}\sigma_{-\mu} \mid p_{1c}^{i}> \hspace{1.15in} \\
-\delta (p_{1}^{f},p_{1}^{i}) \delta (p_{2}^{f},p_{3}^{i}) <n_{2c}^{f} \mid t_{\pm}\sigma_{-\mu} \mid p_{2c}^{i}> \\
+\delta (p_{1}^{f},p_{1}^{i}) \delta (p_{2}^{f},p_{2}^{i}) <n_{2c}^{f} \mid t_{\pm}\sigma_{-\mu} \mid p_{3c}^{i}>] \\
(32) 
\end{eqnarray*}
\begin{eqnarray*}
<p_{1}^{f}p_{2}^{f}p_{3}^{f}p_{4c}^{f} \mid t_{\pm}\sigma_{\mu} \mid p_{1}^{i}p_{2}^{i}p_{3}^{i}n_{1c}^{i}>\hspace{0.3in} \\
=-\delta (p_{2}^{f},p_{1}^{i}) \delta (p_{3}^{f},p_{2}^{i})\delta (p_{4}^{f},p_{3}^{i})\hspace{0.55in} \\
 <p_{1c}^{f} \mid t_{\pm}\sigma_{\mu} \mid n_{1c}^{i}> \hspace{1.1in} \\
 +\delta (p_{1}^{f},p_{1}^{i}) \delta (p_{3}^{f},p_{2}^{i}) \delta (p_{4}^{f},p_{3}^{i}) \hspace{0.7in} \\
 <p_{2c}^{f} \mid t_{\pm}\sigma_{\mu} \mid n_{1c}^{i}> \hspace{1.1in}\\
 -\delta (p_{1}^{f},p_{1}^{i}) \delta (p_{2}^{f},p_{2}^{i}) \delta (p_{4}^{f},p_{3}^{i}) \hspace{0.7in} \\
 <p_{3c}^{f} \mid t_{\pm}\sigma_{\mu} \mid n_{1c}^{i}> \hspace{1.1in} \\
+\delta (p_{1}^{f},p_{1}^{i}) \delta (p_{2}^{f},p_{2}^{i}) \delta (p_{3}^{f},p_{3}^{i}) \hspace{0.7in} \\
 <p_{4c}^{f} \mid t_{\pm}\sigma_{\mu} \mid n_{1c}^{i}> \hspace{1.1in} \\
(33) 
\end{eqnarray*}
\begin{eqnarray*}
<p_{1}^{f}p_{2}^{f}n_{1}^{f}n_{2c}^{f} \mid t_{\pm}\sigma_{\mu} \mid p_{1}^{i}n_{1}^{i}n_{2}^{i}n_{3c}^{i}> \hspace{0.3in} \\
=\delta (p_{1}^{f},p_{1}^{i})[ \delta (n_{1}^{f},n_{2}^{i})\delta (n_{2}^{f},n_{3}^{i}) \hspace{0.65in} \\
<p_{2c}^{f} \mid t_{\pm}\sigma_{\mu} \mid n_{1c}^{i}> \hspace{1.20in} \\
-\delta (n_{1}^{f},n_{1}^{i}) \delta (n_{2}^{f},n_{3}^{i}) <p_{2c}^{f} \mid t_{\pm}\sigma_{\mu} \mid n_{2c}^{i}> \hspace{0.0in} \\
+\delta (n_{1}^{f},n_{1}^{i}) \delta (n_{2}^{f},n_{2}^{i}) <p_{2c}^{f} \mid t_{\pm}\sigma_{\mu} \mid n_{3c}^{i}>] \hspace{0.0in} \\
-\delta (p_{2}^{f},p_{1}^{i})[ \delta (n_{1}^{f},n_{2}^{i})\delta (n_{2}^{f},n_{3}^{i}) \hspace{0.72in} \\
<p_{1c}^{f} \mid t_{\pm}\sigma_{\mu} \mid n_{1c}^{i}> \hspace{1.20in} \\
-\delta (n_{1}^{f},n_{1}^{i}) \delta (n_{2}^{f},n_{3}^{i}) <p_{1c}^{f} \mid t_{\pm}\sigma_{\mu} \mid n_{2c}^{i}> \hspace{0.0in} \\
+\delta (n_{1}^{f},n_{1}^{i}) \delta (n_{2}^{f},n_{2}^{i}) <p_{1c}^{f} \mid t_{\pm}\sigma_{\mu} \mid n_{3c}^{i}>] \hspace{0.0in} \\
(34) 
\end{eqnarray*}
\begin{eqnarray*}
<n_{1}^{f}n_{2}^{f}n_{3}^{f}n_{4c}^{f} \mid t_{\pm}\sigma_{-\mu} \mid p_{1}^{i}n_{1}^{i}n_{2}^{i}n_{3c}^{i}> \hspace{0.15in} \\
= +\delta (n_{2}^{f},n_{1}^{i}) \delta (n_{3}^{f},n_{2}^{i})\delta (n_{4}^{f},n_{3}^{i}) \hspace{0.5in} \\
<n_{1c}^{f} \mid t_{\pm}\sigma_{-\mu} \mid p_{1c}^{i}> \hspace{1.0in} \\
-\delta (n_{1}^{f},n_{1}^{i}) \delta (n_{3}^{f},n_{2}^{i}) \delta (n_{4}^{f},n_{3}^{i}) \hspace{0.65in} \\
<n_{2c}^{f} \mid t_{\pm}\sigma_{-\mu} \mid p_{1c}^{i}> \hspace{1.0in}\\
+\delta (n_{1}^{f},n_{1}^{i}) \delta (n_{2}^{f},n_{2}^{i}) \delta (n_{4}^{f},n_{3}^{i}) \hspace{0.65in} \\
<n_{3c}^{f} \mid t_{\pm}\sigma_{-\mu} \mid p_{1c}^{i}> \hspace{1.0in} \\
-\delta (n_{1}^{f},n_{1}^{i}) \delta (n_{2}^{f},n_{2}^{i}) \delta (n_{3}^{f},n_{3}^{i}) \hspace{0.65in} \\
<n_{4c}^{f} \mid t_{\pm}\sigma_{-\mu} \mid p_{1c}^{i}> \hspace{1.0in} \\
(35) 
\end{eqnarray*}
For all the given q.p. transition amplitudes [Eqs. (22)- (35)], we took into account the antisymmetrization of the single- q.p. states:\\
$ p_{1}^{f}<p_{2}^{f}<p_{3}^{f}<p_{4}^{f}$,\\
$ n_{1}^{f}<n_{2}^{f}<n_{3}^{f}<n_{4}^{f}$,\\
$ p_{1}^{i}<p_{2}^{i}<p_{3}^{i}<p_{4}^{i}$,\\
$ n_{1}^{i}<n_{2}^{i}<n_{3}^{i}<n_{4}^{i}$.\\
GT transitions of phonon excitations for every excited state were also taken into account. Here we assumed that the q.p.'s in the parent nucleus remained in the same q.p. orbits. A detailed description of the formalism can be found in [25].
\subsection{Application of pn-QRPA to Calculate Nuclear Matrix Elements}

Construction of parent excited states and the corresponding transition amplitudes to daughter excited states are based on the formalism discussed in subsection (3.1). The effects of $S_{p}$ and $S_{n}$ of a nucleus were considered for the first time in the calculation of the stellar weak rates. For each nucleus we derived the corresponding $S_{p}$ and $S_{n}$ from mass tables. The minimum of $S_{p}$ and $S_{n}$ was considered as a  cut-off excitation energy for the parent nucleus. Due to uncertainties in the calculation of energy levels and the effect of Coulomb barrier for the case of proton emission, we set our cut-off excitation energies 1 MeV higher than the minimum of $S_{p}$ and $S_{n}$ for parent nuclei.

As can be seen from the transition amplitudes [Eqs. (24)-(29), (32-35)], excited states of even-odd and odd-odd nuclei are also constructed from three- and four-q.p. states, respectively. Hence the number of our initial states are sometimes large compared to OMHTS for our set of cut-off energies (eg., we considered 183 initial states for the nucleus $^{\mathrm{36}}$Cl, 160 initial states for the nucleus $^{\mathrm{32}}$P, and so on). OHMTS considered only the lowest 100 initial excited states or all states when they were less than 100. Our computer program solves the RPA equation for every single state keeping in view the blocking effect [29]. For each nucleus the first 100 final states were considered. We assured satisfactory convergence in our rates. 

In the calculation of GT transitions we did not explicitly take into account the quenching of the transitions. One reason is that the quenching of the GT strength cannot be a constant renormalization of the axial vector current [30,31]. An extra quenching factor on one hand might slightly improve the experimental half-lives for some $\beta^{+}$ decays of very proton-rich $\mathit{sd}$- shell nuclei but the overall comparison with measured rates of $\mathit{sd}$- shell nuclei would remain more or less unaltered (see [6]).

As far as the superallowed Fermi transitions are concerned, we assume them to be concentrated in the isobaric analogue state (IAS) of the parent state. The Fermi matrix element depends only on the nuclear isospin, T, and it's projection $T_{z}=(Z-N)/2$ for the parent and daughter nucleus. The energy of the IAS is calculated according to the prescription given in [32, page 111-112]
whereas the reduced transition probability is given by 
\begin{displaymath}
B(F)=T(T+1)-T_{zi}T_{zf} \hspace{0.3in} (36) 
\end{displaymath}
where $T_{zi}$ and $T_{zf}$ are the third components of the isospin of initial and final analogue states.

\section{\normalsize INCORPORATION OF EXPERIMENTAL DATA}
 Since the work of OHMTS, new experimental data for $\mathit{sd}$- shell nuclei have been compiled. The latest experimental excitation energies and log $\mathit{ft}$ values are taken from updated compilations by Tilley et al. [33] for A=17, [34] for A=18-19, Ajzenberg-Selove [35] for A=20, and, Endt [36] for A=21-39. We augmented all relevant experimental data in our calculations.

We replaced our calculated excitation energies with the measured ones when they were within 0.5 MeV of each other. The log $\mathit{ft}$ value of this energy level was also then replaced by the measured one. Very low lying states were inserted in our calculations together with their log $\mathit{ft}$ values if our theory was missing them. Inverse and mirror transitions were also taken into consideration. If there appeared a level in experimental compilations without definite spin and parity assignment, we did not replace theoretical levels with experimental levels beyond this excitation energy nor did we insert it.

The Q-value of each transition as well as $S_{p}$ and $S_{n}$ of each nucleus were calculated using the experimental mass compilation of Audi et al. [37]. For nuclei where [37] failed to give mass defects we used M\"oller and Nix [38] to derive the corresponding energies necessary for our calculations.

\section{\normalsize RESULTS AND DISCUSSIONS}
\subsection{Charge-changing Transition Matrix Elements}
Eq. ~(1) can be factorized into three parts,
\begin{displaymath}
\lambda_{ij}= \frac{ln2}{D} f_{ij} B_{ij} \\
\hspace{1.25in} (37)
\end{displaymath}
The first factor is a constant and the second are the phase space integrals, which can be calculated relatively accurate. So reliable rates depend directly on reliable calculation of matrix elements. Our work provides a new option for the calculation of $B_{ij}$. Again as can be seen from Eq. ~(4), reliable calculation of $B_{ij}$ implies reliable calculation of GT transitions, since superallowed Fermi transitions can easily be calculated from Eq. ~(37). The GT transitions contain the nuclear structure details of the parent and daughter nuclei. 

The distribution of the strength function in the pn-QRPA model has already been discussed extensively, and as such we will not be discussing it here. Interested readers are referred to, eg., [5,8]. The inclusion of particle-particle forces leads to enhancement of the ground state correlations and a redistribution of the calculated beta strength [8,28].

The terrestrial half-life calculations using pn-QRPA are in good agreement with measured values. A good comparison of our theory with measured data for neutron-rich nuclei is given in [6]. In Fig. 8 of [6] a direct comparison of pn-QRPA theory and the shell-model calculation with experimental data is given. $\beta^{+}/EC$ decay half-lives are compared with the corresponding experimental values in [8]. In both cases the comparison is fairly good, encouraging us to apply this theory also to transitions from excited states.
\subsection{Comparison with Earlier Work }
\subsubsection{OHMTS}
Prior to our calculations, OHMTS and FFN performed an extensive calculation of the rates of $\mathit{sd}$- shell nuclei. Our work and the compilation of OHMTS are tabulated at the same grid points of density and temperature, for the ranges  (10$\leq$ $\mathit{\rho Y_{e}}$ (gcm$^{-3}$) $ \leq$ 10$^{11}$, and 10$^{7}$ $\leq$ T(K) $\leq$ 30 x 10$^{9}$). The effect of particle emission was totally ignored in previous calculations. For example, consider the $\beta^{-}$ decay of $^{\mathrm{32}}$Cl by OHMTS. The Q-value of this decay is -11.158 MeV, and the $S_{p}$ of this nucleus is 1.574 MeV. At 10$^{7}$K OHMTS give a $\beta^{-}$ decay rate of 10$^{-100}$, i.e. zero. For higher temperatures, however, OHMTS $\beta^{-}$ decay rates are finite, implying that the parent nucleus must be in an excited state with an excitation energy of at least 11.158 MeV. But, physically, far below this energy, proton emissions will start from this nucleus (see also table 32.27, page 393 of [36]), and hence the $\beta^{-}$ decay rate should be zero. FFN also got $\beta^{-}$ decay rates for this isotope of chlorine at higher temperatures. We believe that this description does not correspond to reality, and there exists no $\beta^{-}$ decay rate for this isotope. $^{\mathrm{32}}$Cl can however decay to $^{\mathrm{32}}$Ar via capturing a positron. The phase space integrals for capture rates are non-zero even for negative Q-values [Eq. ~(8)] and hence we do get positron capture rates around 1.5 MeV for $^{\mathrm{32}}$Cl. There are many more such cases where the Q-value of the decay reaction is a big negative number and the absolute Q-value of this reaction is much larger than the minimum of $S_{p}$ and $S_{n}$. In all such cases there could be no decay rate possible and the nucleus will rather emit protons or neutrons (as the case may be). 

In our QRPA calculation, we considered a very large model space including excitations up to 7 major shells (7$\hbar \omega$ of energy). One of the selection rules for the quasi-particle transition to occur is that the transforming quasi-particles should belong to the same major oscillator shell, i.e the neutron which changes to a proton should belong to the same major oscillator shell (and same for protons). There are six cases of $\beta^{-}$ decay rates ($^{22}$Na, $^{26}$Al, $^{29}$Si, $^{31}$P, $^{33}$S, and  $^{37}$Ar) and ten cases of $\beta^{+}$ decay rates ($^{21}$Ne, $^{24}$Na, $^{28}$Al, $^{29}$Al, $^{29}$Si, $^{30}$Si, $^{32}$P, $^{33}$P, $^{34}$P, and $^{35}$S) where the decay channel opens below the particle decay channel but we do not obtain any decay rates. In all these cases the above mentioned selection rule is satisfied only for high-lying major shells but then the parent excited state (which is the sum of the energies of constituent quasi-particles) is above the particle decay energy. 

In our table of weak rates, we see that for most cases either the  $\beta^{-}$ or $\beta^{+}$ decay rates are zero (depending on the Q-values). Only for 10 of the 90 $\mathit{sd}$- shell nuclei considered ($^{23}$Na, $^{25}$Na, $^{25}$Mg, $^{26}$Mg, $^{27}$Al, $^{34}$S, $^{35}$Cl, $^{36}$Cl, $^{37}$Cl, and $^{38}$Ar), we get both $\beta^{-}$ and $\beta^{+}$ decay rates using the pn-QRPA theory. In all cases with negative Q values (except the above mentioned 16 cases) the particle decay channel opens below the $\beta$ decay channel and hence we do not obtain any decay rates. This is in sharp contrast to the earlier works of FFN and OHMTS.

Table A depicts the nuclei which were considered in the calculations of weak rates by us and by OHMTS. OHMTS calculated four different weak rates (capture rates, decay rates, gamma heating rates and (anti)neutrino energy loss rates) for 158 nuclei. We calculated six different weak rates (the above mentioned four rates plus energy rates of particle emission processes and probability of these $\beta$-delayed emission processes) for 178 nuclei.

Our cut-off excitation energies, in parent nuclei, are constrained due to the effect of  $S_{p}$ and $S_{n}$. This picture, when compared to ignoring the particle emission processes totally, is closer to reality, and generally our cut-off excitation energies are lower compared to OHMTS. At higher temperatures, where the occupation probability of excited states is bigger, our rates are then suppressed compared to OHMTS. Often our theory gives high-lying transitions tending to enhance our rates as compared to OHMTS. Our rates are a competitive interplay of these two effects.

At the oxygen-burning stage of heavy stars (20 - 25 M$_{\odot}$), the electron capture rate of $^{\mathrm{33}}$S and $^{\mathrm{35}}$Cl plays a decisive role. Figure~1 shows the comparison of our electron capture rates with those of OHMTS for the nucleus $^{\mathrm{33}}$S.  
\begin{figure}
\epsfxsize=7.8cm
\epsffile{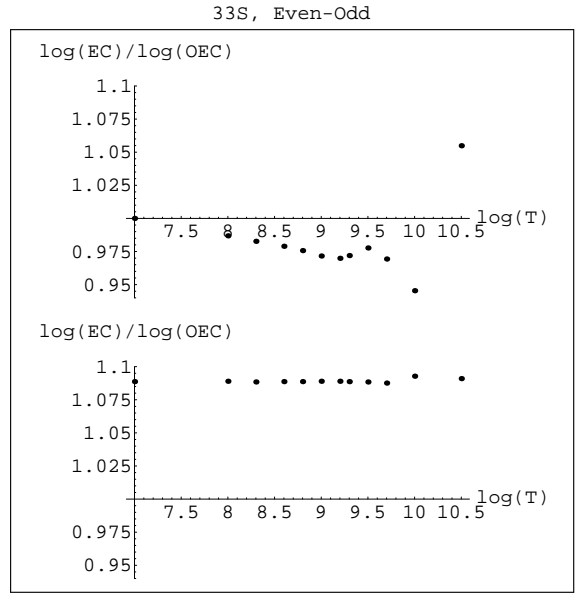}
\caption{ \footnotesize Comparison of our log rates of electron capture (EC) with that of OHMTS (OEC). Log(T) is the log of temperature in units of Kelvin. The upper graph is plotted at $\rho Y_{e}=$ 10 g cm$^{-3}$, and the lower graph is plotted at $\rho Y_{e}=$ 10$^{11}$ g cm$^{-3}$.}
\end{figure}
In Fig. 1, EC denotes our electron capture rates, and OEC denotes the electron capture rates of OHMTS. The upper graph is at density $\rho Y_{e}=$ 10 g cm$^{-3}$, while the lower graph is at density $\rho Y_{e}=$ $10^{11}$ g cm$^{-3}$, the two extreme limits of the density scale. As can be seen, the comparison is fair at both density scales. Figure~2 shows the comparison for the case of $^{\mathrm{35}}$Cl while Figure~3 depicts the situation for $^{\mathrm{27}}$Al.
\begin{figure}
\epsfxsize=7.8cm
\epsffile{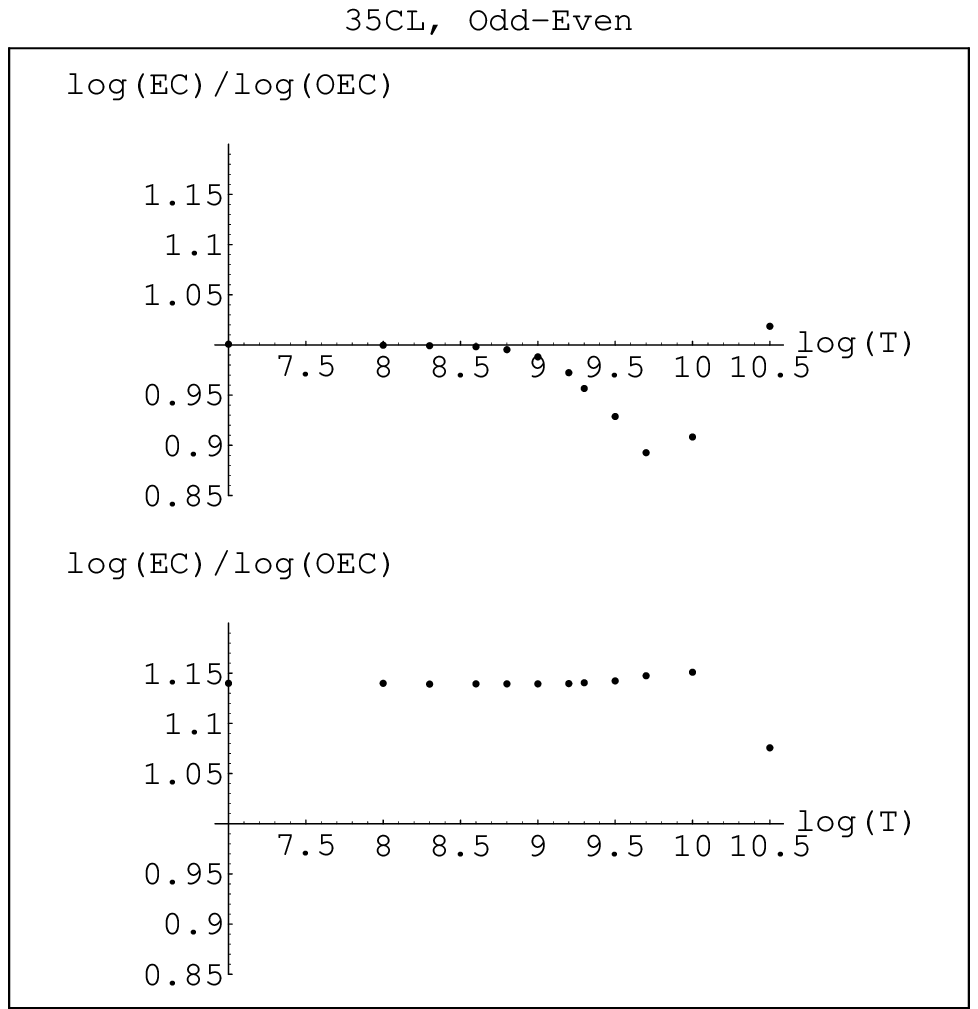}
\caption{ \footnotesize Comparison of our log rates of electron capture (EC) with that of OHMTS (OEC). Log(T) is the log of temperature in units of Kelvin. The upper graph is plotted at $\rho Y_{e}=$ 10 g cm$^{-3}$, and the lower graph is plotted at $\rho Y_{e}=$ 10$^{11}$ g cm$^{-3}$.}
\end{figure}
\begin{figure}
\epsfxsize=7.8cm
\epsffile{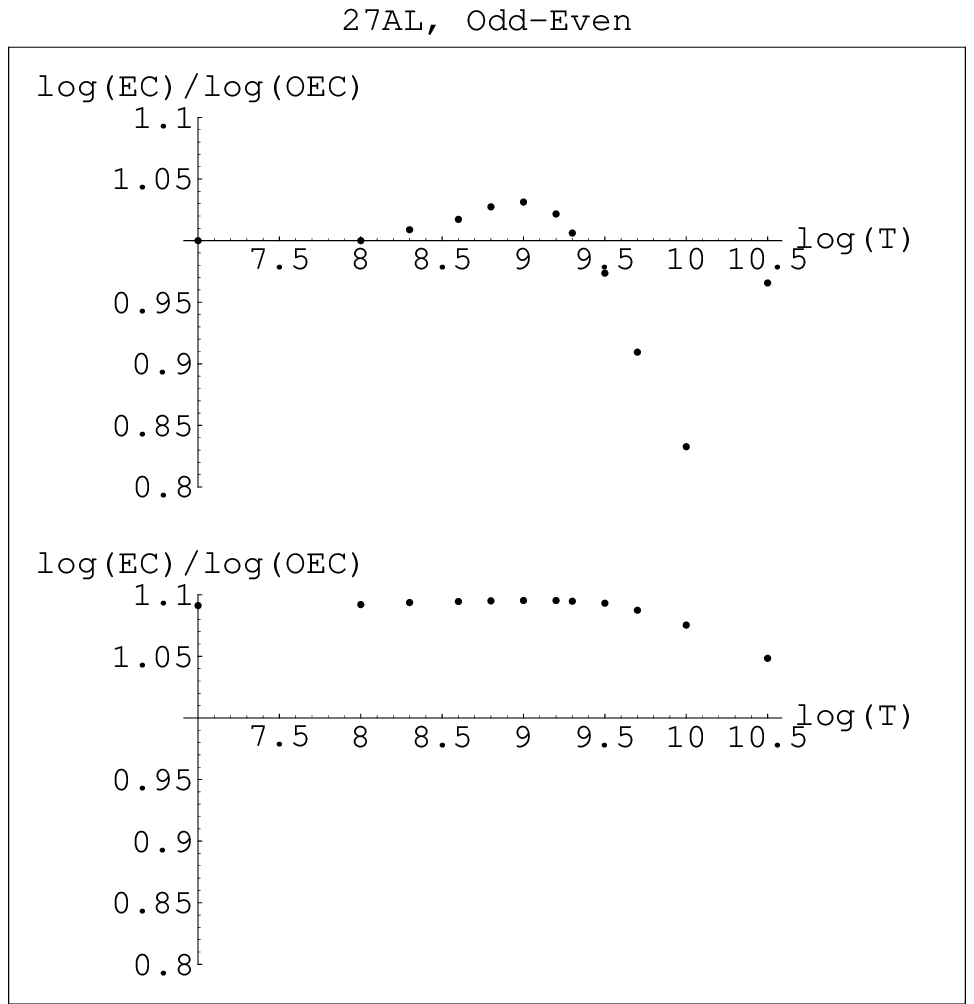}
\caption{ \footnotesize Comparison of our log rates of electron capture (EC) with that of OHMTS (OEC). Log(T) is the log of temperature in units of Kelvin. The upper graph is plotted at $\rho Y_{e}=$ 10 g cm$^{-3}$, and the lower graph is plotted at $\rho Y_{e}=$ 10$^{11}$ g cm$^{-3}$.}
\end{figure}
It is clear from these figures that our rates are suppressed at higher temperatures and densities. We are in fair agreement at low temperatures and densities.

Often our theory gives higher lying transitions compared to the shell model, and then at higher temperatures (9 $\leq$ log(T) $\leq$ 10) our rates are enhanced compared to those of OHMTS. Figure~4 and Figure~5 depicts two such cases. Fig.~4 shows the even-even nucleus $^{\mathrm{20}}$Ne while the odd-odd nucleus $^{\mathrm{22}}$Na is depicted in Fig. 5. The interplay of the two effects, discussed earlier, is evident in these figures.
\begin{figure}
\epsfxsize=7.8cm
\epsffile{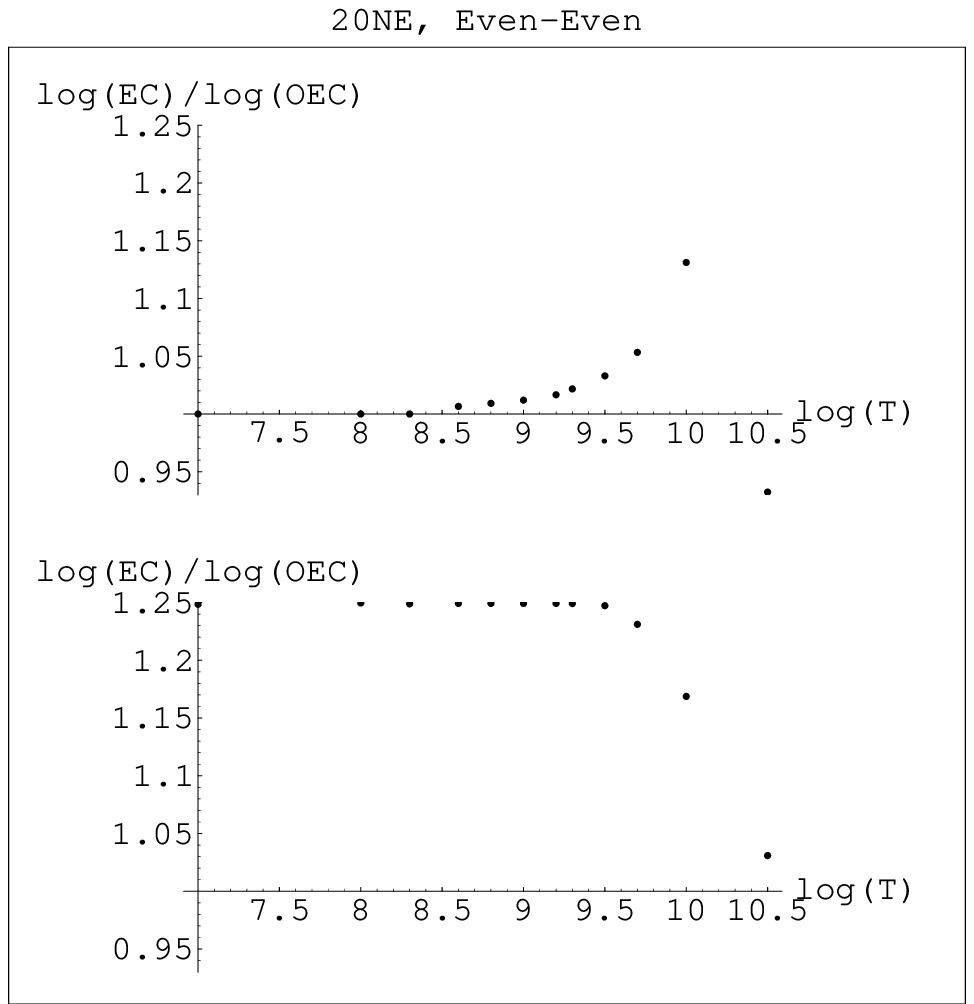}
\caption{ \footnotesize Comparison of our log rates of electron capture (EC) with that of OHMTS (OEC). Log(T) is the log of temperature in units of Kelvin. The upper graph is plotted at $\rho Y_{e}=$ 10 g cm$^{-3}$, and the lower graph is plotted at $\rho Y_{e}=$ 10$^{11}$ g cm$^{-3}$.}
\end{figure}
\begin{figure}
\epsfxsize=7.8cm
\epsffile{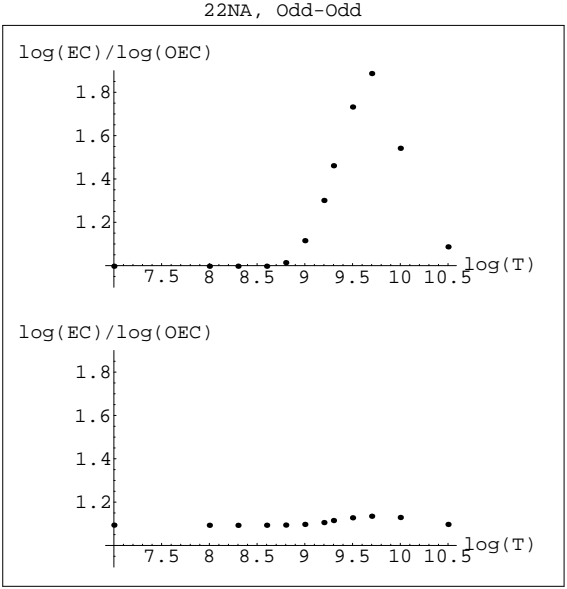}
\caption{ \footnotesize Comparison of our log rates of electron capture (EC) with that of OHMTS (OEC). Log(T) is the log of temperature in units of Kelvin. The upper graph is plotted at $\rho Y_{e}=$ 10 g cm$^{-3}$, and the lower graph is plotted at $\rho Y_{e}=$ 10$^{11}$ g cm$^{-3}$.}
\end{figure}

Next we present some comparisons of the positron decay rates. The positron decay rates of $^{\mathrm{30}}$P, $^{\mathrm{31}}$S, and $^{\mathrm{34}}$Cl are important at the oxygen-burning stage of heavy stars (20 - 25 M$_{\odot}$). Figure~6 shows a comparison between the positron decay rates for the case of $^{\mathrm{30}}$P.
\begin{figure}
\epsfxsize=7.8cm
\epsffile{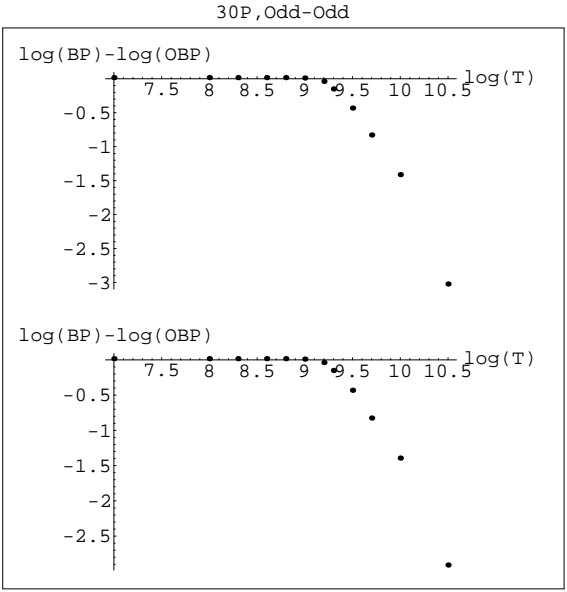}
\caption{ \footnotesize Comparison of our log rates of positron decay (BP) with that of OHMTS (OBP). Log(T) is the log of temperature in units of Kelvin. The upper graph is plotted at $\rho Y_{e}=$ 10 g cm$^{-3}$, and the lower graph is plotted at $\rho Y_{e}=$ 10$^{11}$ g cm$^{-3}$.}
\end{figure}
In Fig. 6, BP denotes our positron decay rates and OBP denotes positron decay rates of OHMTS. Also note that here the differences of the two log rates are plotted as a function of temperature. Previously, for the comparison of electron capture rates, we chose the ratio of the two log rates. We decided to consider differences of log rates here because at certain temperatures the log rates of OHMTS were approaching zero, causing an abnormal increment in the ratio of the two rates. In comparing the positron decay rates, we again depict two graphs in each figure. The upper graph is at density $\rho Y_{e}=$ 10 g cm$^{-3}$, while the lower graph is at density $\rho Y_{e}=$ $10^{11}$ g cm$^{-3}$. We see that the agreement is perfect at lower temperatures. As temperature increases, our rates are again suppressed compared to those of OHMTS. OHMTS also consider transitions from excited states much higher than the minimum of $S_{p}$ and $S_{n}$ value which causes their rates to increase at higher temperatures. We argue that such contributions should not be considered. At any $E_{i}$, higher than the minimum of $S_{p}$ and $S_{n}$ (after accounting for the effective Coulomb barrier which prevents a proton from being promptly emitted, and, the uncertainty in calculation of energy levels), the nucleus will emit protons or neutrons (as the case may be) instead of continuing to undergo $\beta^{+}$ decay. Figure~7 and Figure~8 show a similar situation for the $\beta^{+}$ decay of $^{\mathrm{31}}$S and $^{\mathrm{21}}$Na, respectively. Again the differences in rates increases with increasing temperature (at $\rho Y_{e}=$ 10 g cm$^{-3}$ and log T= 10.5, our rates are suppressed by more than a factor of 27 for the case of $^{\mathrm{31}}$S while for $^{\mathrm{21}}$Na our rates are suppressed by more than two orders of magnitude).  
\begin{figure}
\epsfxsize=7.8cm
\epsffile{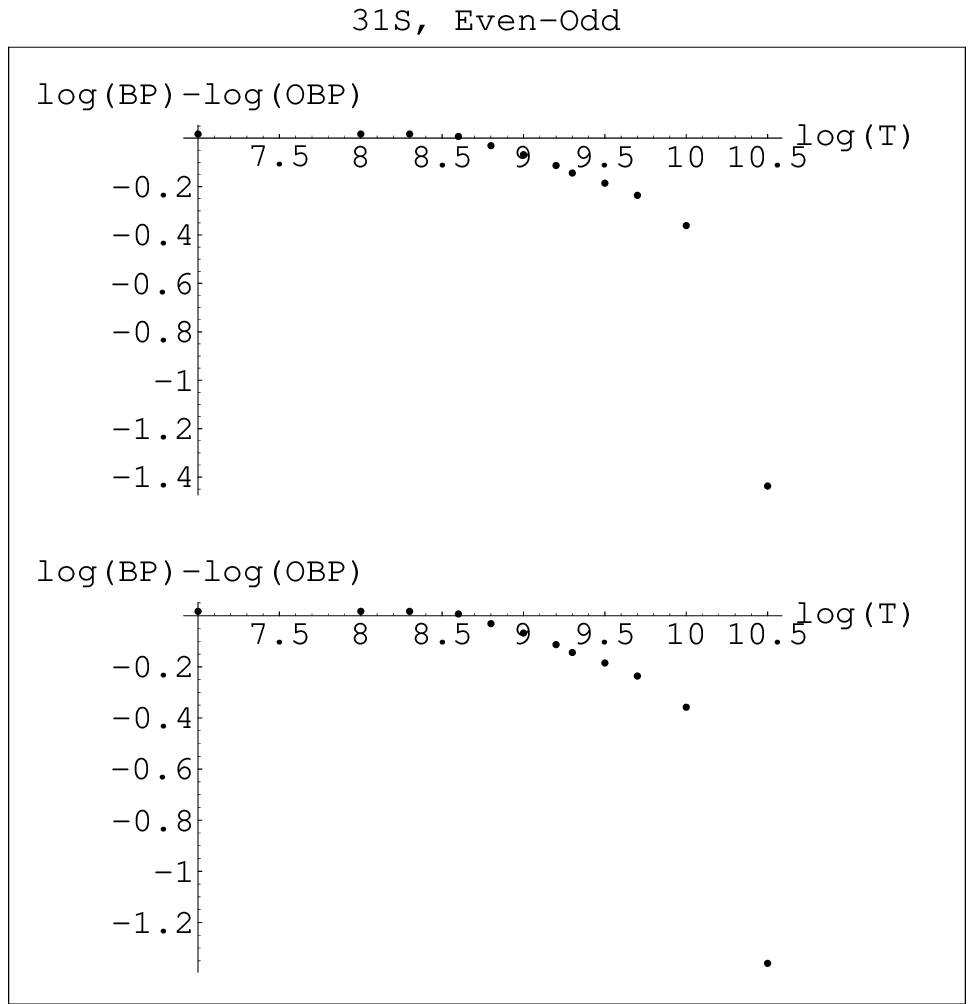}
\caption{ \footnotesize Comparison of our log rates of positron decay (BP) with that of OHMTS (OBP). Log(T) is the log of temperature in units of Kelvin. The upper graph is plotted at $\rho Y_{e}=$ 10 g cm$^{-3}$, and the lower graph is plotted at $\rho Y_{e}=$ 10$^{11}$ g cm$^{-3}$.}
\end{figure}
\begin{figure}
\epsfxsize=7.8cm
\epsffile{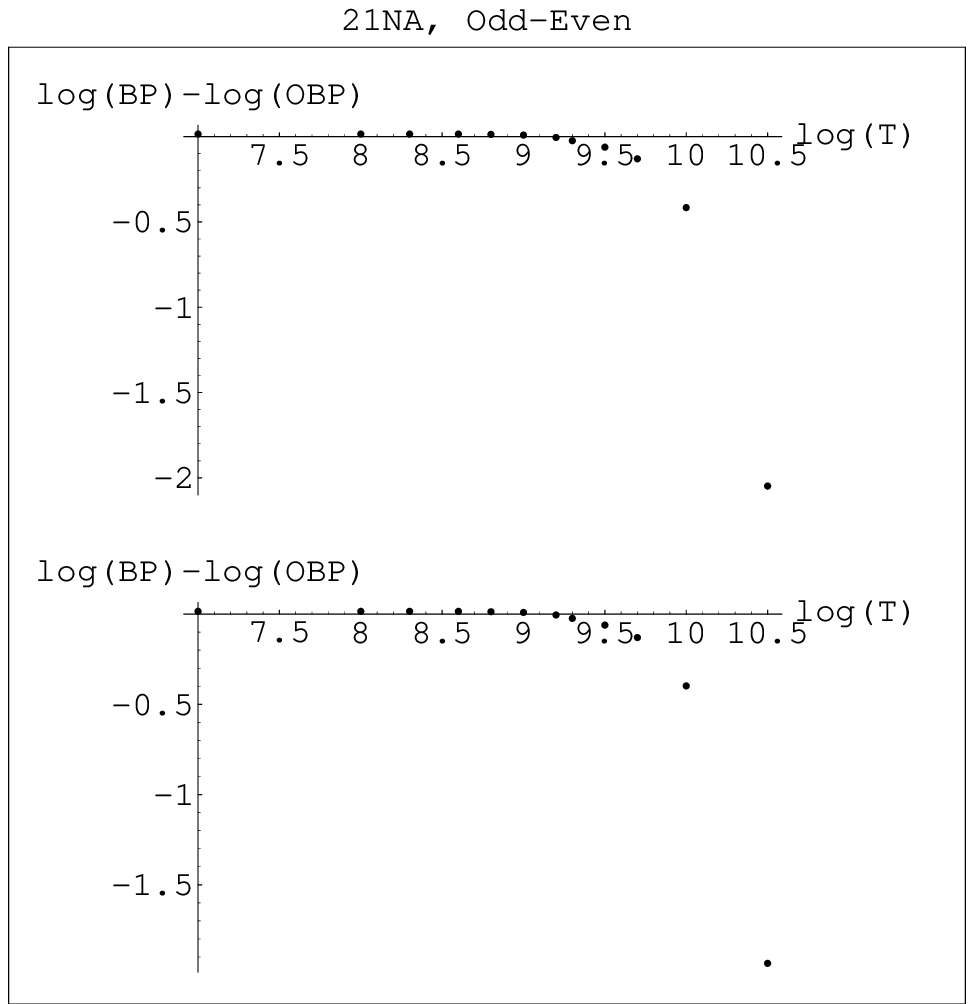}
\caption{ \footnotesize Comparison of our log rates of positron decay (BP) with that of OHMTS (OBP). Log(T) is the log of temperature in units of Kelvin. The upper graph is plotted at $\rho Y_{e}=$ 10 g cm$^{-3}$, and the lower graph is plotted at $\rho Y_{e}=$ 10$^{11}$ g cm$^{-3}$.}
\end{figure}

As example of the cases where pn-QRPA theory gives higher lying transitions compared to the shell model, we present the positron decay rates of $^{\mathrm{34}}$Cl and $^{\mathrm{26}}$Si. Fig.~9 shows the comparison of positron decay rates for the nucleus $^{\mathrm{34}}$Cl. One should also note that for the above mentioned cases of $^{\mathrm{30}}$P, $^{\mathrm{31}}$S, $^{\mathrm{21}}$Na, and $^{\mathrm{34}}$Cl our $\beta^{+}$ rate [Eq. ~(15)] decreases as the temperature increases. This is not a surprise if we consider Eq. ~(14). From there we see the probability of occupation of low-lying states decrease as temperature increases (as now there is also a finite probability of occupation of high-lying states). As indicated earlier, in pn-QRPA theory, the decay transition is possible if the contributing quasi-particles belong to the same major oscillator shell. These high-lying states have no $\beta^{+}$ decay transitions in our selected examples, and hence our overall rate decreases at higher temperatures. This is not the case, for example, in $^{\mathrm{26}}$Si. Here we do get $\beta^{+}$ decay transitions also at high excitation energies. So we see that the decay rate increases with increasing temperature. Figure~10 shows a comparison of the $\beta^{+}$ decay rates for the even-even case of $^{\mathrm{26}}$Si. 
\begin{figure}
\epsfxsize=7.8cm
\epsffile{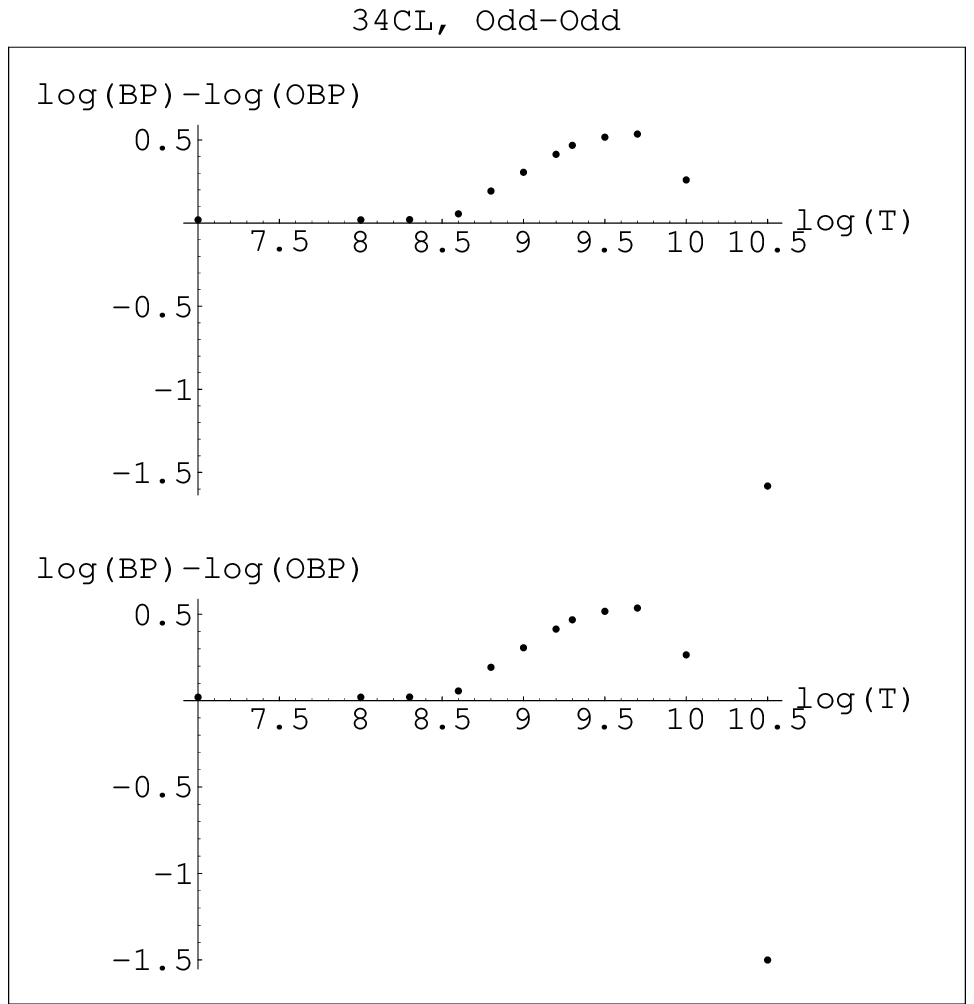}
\caption{ \footnotesize Comparison of our log rates of positron decay (BP) with that of OHMTS (OBP). Log(T) is the log of temperature in units of Kelvin. The upper graph is plotted at $\rho Y_{e}=$ 10 g cm$^{-3}$, and the lower graph is plotted at $\rho Y_{e}=$ 10$^{11}$ g cm$^{-3}$.}
\end{figure}
\begin{figure}
\epsfxsize=7.8cm
\epsffile{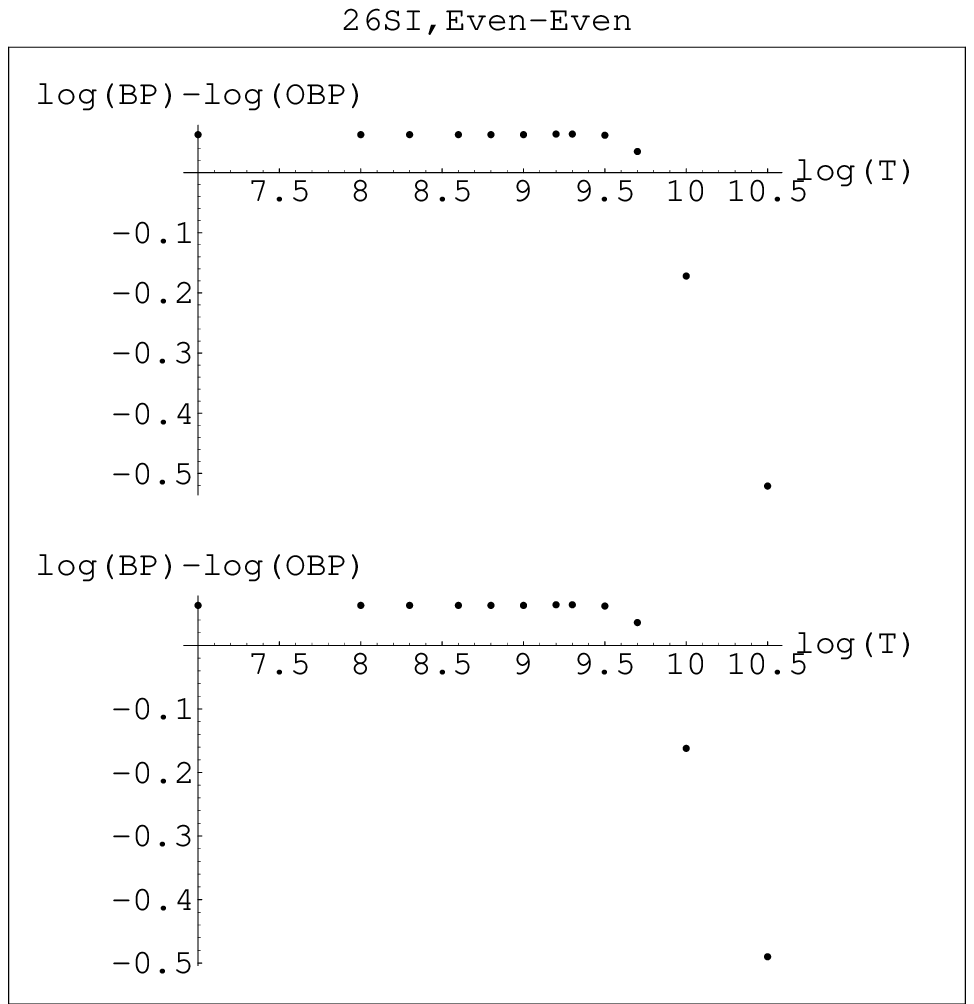}
\caption{ \footnotesize Comparison of our log rates of positron decay (BP) with that of OHMTS (OBP). Log(T) is the log of temperature in units of Kelvin. The upper graph is plotted at $\rho Y_{e}=$ 10 g cm$^{-3}$, and the lower graph is plotted at $\rho Y_{e}=$ 10$^{11}$ g cm$^{-3}$.}
\end{figure}

The comparison of positron capture rates and electron decay rates is similar to the preceding comparison of electron capture rates and positron decay rates. In our table of rates, -100 means that the rate is smaller than $10^{-100}$. It should be pointed out that the decay rates are sensitive functions of the difference of parent and daughter energies ($E_{i}-E_{j}$) and just by a mere addition of around 0.5 MeV, many of our -100 decay rates could rise to larger numbers. 
\begin{figure}[h!]
\epsfxsize=7.8cm
\epsffile{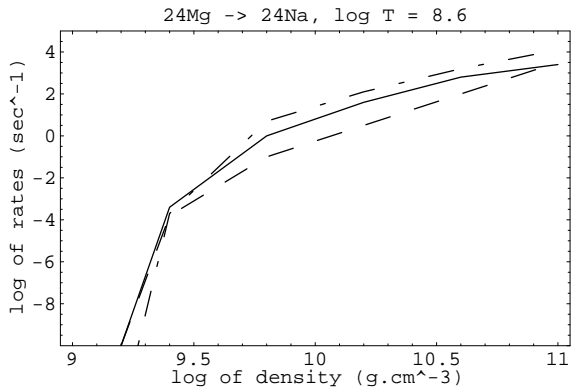}
\caption{ \footnotesize Electron capture rates for $^{\mathrm{24}}$Mg as a function of $\rho Y_{e}$, at log T = 8.6. The dot-dashed line represents our rates, the dashed line depicts the rates of FFN, while the solid line shows the results of OHMTS.}
\end{figure}
\begin{figure}[h!]
\epsfxsize=7.8cm
\epsffile{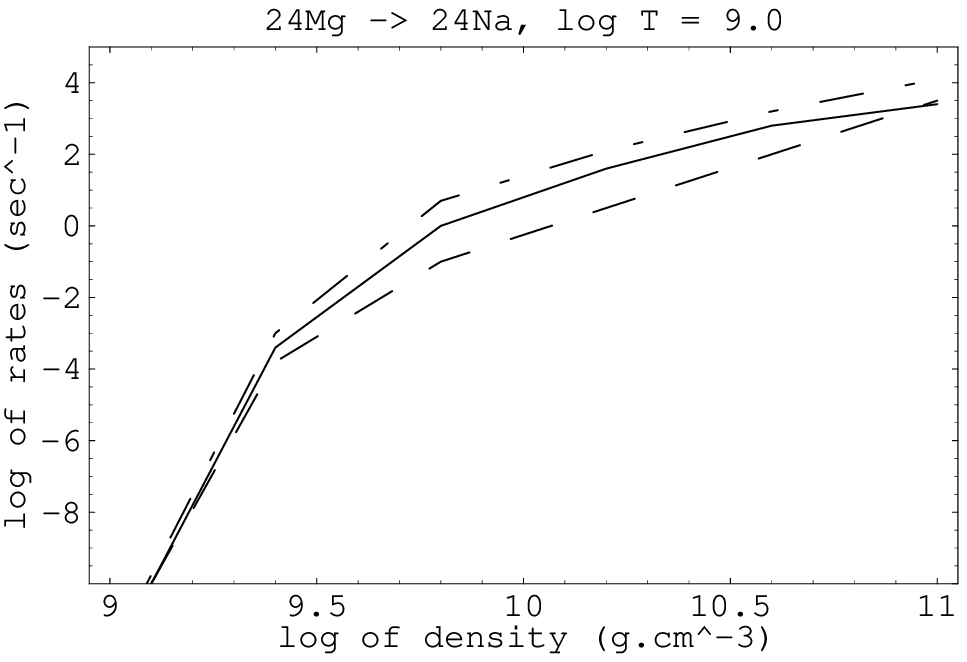}
\caption{ \footnotesize Electron capture rates for $^{\mathrm{24}}$Mg as a function of $\rho Y_{e}$, at log T = 9.0. The dot-dashed line represents our rates, the dashed line depicts the rates of FFN, while the solid line shows the results of OHMTS.}
\end{figure}
\begin{figure}
\epsfxsize=7.8cm
\epsffile{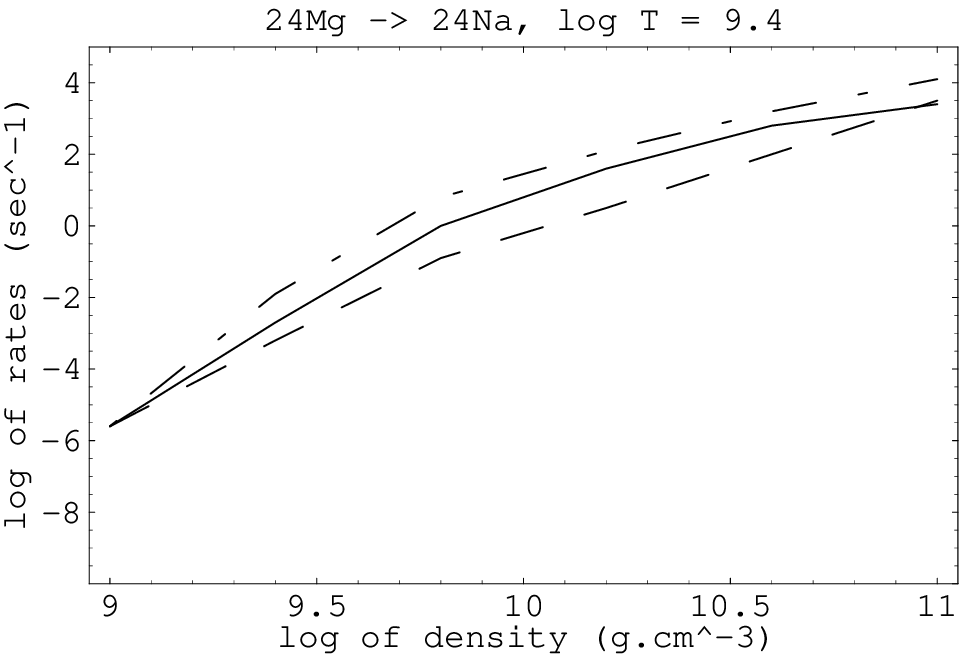}
\caption{ \footnotesize Electron capture rates for $^{\mathrm{24}}$Mg as a function of $\rho Y_{e}$, at log T = 9.4. The dot-dashed line represents our rates, the dashed line depicts the rates of FFN, while the solid line shows the results of OHMTS.}
\end{figure}
\begin{figure}[h!]
\epsfxsize=7.8cm
\epsffile{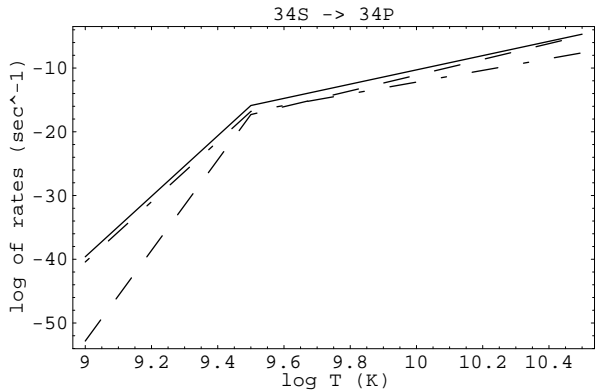}
\caption{ \footnotesize Electron decay rates for $^{\mathrm{34}}$S as a function of temperature, at $\rho Y_{e}=$ 10$^{3}$ g cm$^{-3}$. The dot-dashed line represents our rates, the dashed line depicts the rates of FFN, while the solid line shows the results of OHMTS.}
\end{figure}
\subsubsection{OHMTS and FFN}
For all nuclei, not far from the line of stability, where experimental data are also available, our results, the results of OHMTS, and, the results of FFN are always in good agreement. 
To present a mutual comparison of the three calculations, for such a nuclei, graphically, we selected the electron capture process of $^{\mathrm{24}}$Mg.
 We present the comparison at log T = 8.6, log T = 9.0, and, log T =9.4. Figure~11 shows the comparison at log T = 8.6.
In Fig. 11, our electron capture rates are denoted by the dot-dashed line, the rates of FFN by dashed line, and, the rates of OHMTS are represented by a solid line. We see that the comparison between the different rates is fairly good.
Figure~12 shows the same process at a much higher temperature of log T = 9.0.
Finally we depict the same situation at log T = 9.4 in Fig.~13.

As stated earlier, the incorporation of experimental data tends to give similar results. Only for those nuclei where measured transitions are not available we begin to differ in weak rates. The underlying formalism then plays the key role in calculating the transitions. We show here typical cases of electron decay rates. 
 Figure~14 shows the electron decay rates of $^{\mathrm{34}}$S. Here the difference in the rate calculations is more obvious.
\begin{figure}[h!]
\epsfxsize=7.8cm
\epsffile{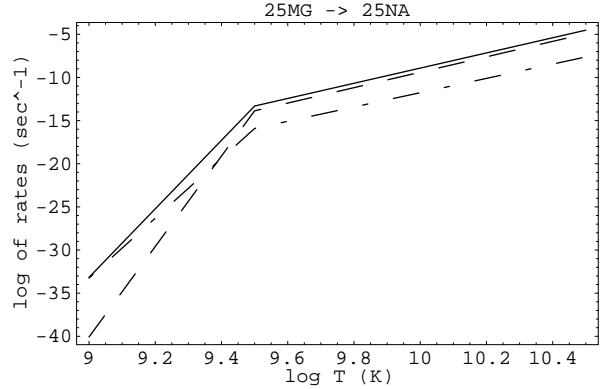}
\caption{ \footnotesize Electron decay rates for $^{\mathrm{25}}$Mg as a function of temperature, at $\rho Y_{e}=$ 10$^{3}$ g cm$^{-3}$. The dot-dashed line represents our rates, the dashed line depicts the rates of FFN, while the solid line shows the results of OHMTS.}
\end{figure}
\begin{figure}[h!]
\epsfxsize=7.8cm
\epsffile{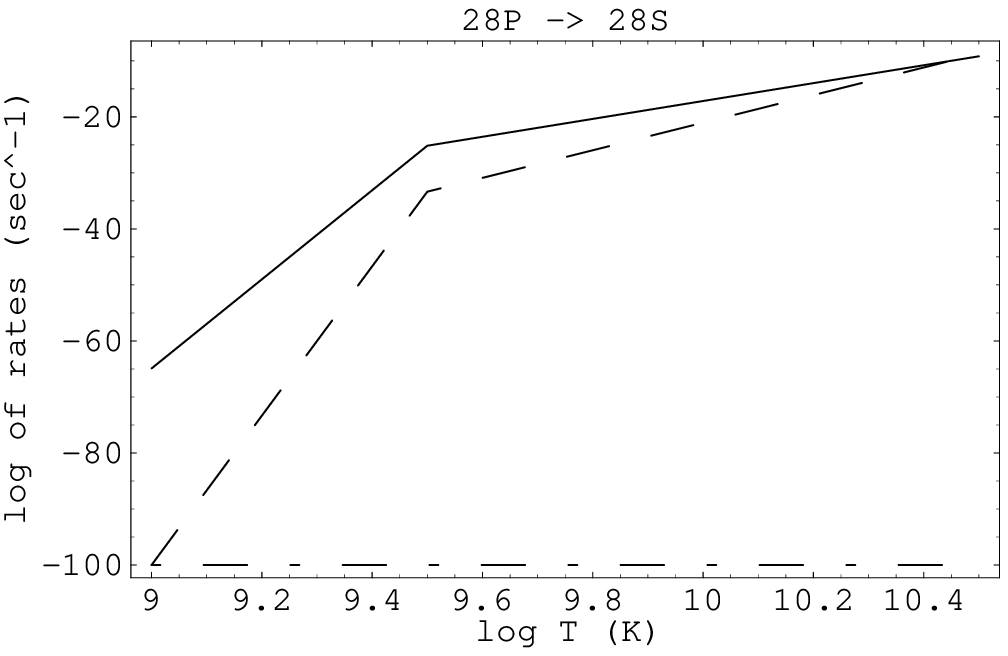}
\caption{ \footnotesize Electron decay rates for $^{\mathrm{28}}$P as a function of temperature, at $\rho Y_{e}=$ 10$^{3}$ g cm$^{-3}$. The dot-dashed line represents our rates, the dashed line depicts the rates of FFN, while the solid line shows the results of OHMTS.}
\end{figure}
\begin{figure}[h!]
\epsfxsize=7.8cm
\epsffile{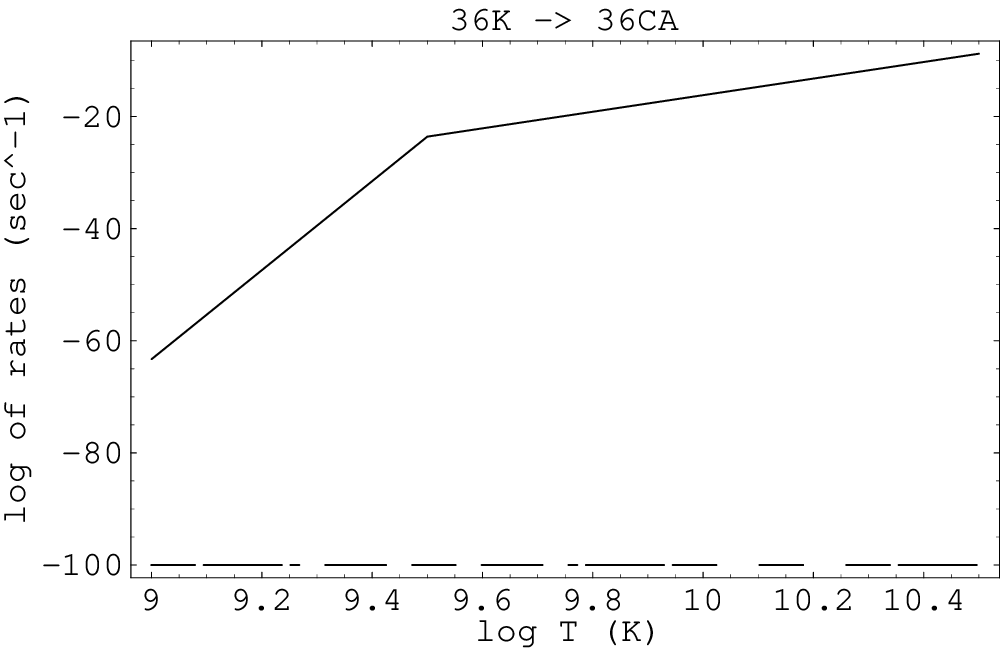}
\caption{ \footnotesize Electron decay rates for $^{\mathrm{36}}$K as a function of temperature, at $\rho Y_{e}=$ 10$^{3}$ g cm$^{-3}$. The dot-dashed line represents our rates, the dashed line depicts the rates of FFN, while the solid line shows the results of OHMTS.}
\end{figure}
In Fig. 15 we depict the electron decay rates of $^{\mathrm{25}}$Mg. Large differences in decay rates are again significant.

In extreme cases for nuclei with no measured transitions, we also encounter situations when OHMTS calculate decay rates for nuclei where we or FFN present no rates. Figure~16 shows the electron decay rates of $^{\mathrm{28}}$P. At temperature 10$^{9}$K, only OHMTS report a decay rate. In Fig~17, which depicts the decay rates of $^{\mathrm{36}}$K, both we and FFN did not find any non-vanishing decay rates.
 
Before concluding this section we would also like to comment on the probabilities of the emitted particles in our calculation.The probabilities of $\beta$-delayed proton (neutron) $P^{p(n)}$ emission at temperature 10$^{7}$ K and density $\rho Y_{e}=$ 10 g cm$^{-3}$ can be compared to terrestrial probabilities. Unfortunately measured $P^{p(n)}$ are not always given. We however searched for these measured probabilities to check the reliability of our calculations. Often it is only stated that particle emission processes are observed. For all such nuclei, we also do calculate finite probabilities $P^{p(n)}$. The agreement with the experimental values is satisfactory (see also the discussion in [39]). 
\section{\normalsize SUMMARY}

The pn-QRPA theory is used for the first time for the calculation of weak interaction rates for $\mathit{sd}$- shell nuclei in stellar environment. The pn-QRPA theory which gives good results for the terrestrial rates also promises to be a good candidate for stellar rates. Our results are compared with that of OHMTS and FFN.

We considered, for the first time, in stellar conditions, the effects of proton and neutron emissions from excited states which affected our decay rates significantly. FFN and OHMTS ignored this effect. Consideration of this effect does lead to the result that many of the beta decay rates are zero when the Q-values are negative. However we believe that this depicts a more realistic picture. The capture rates are always possible and given in our table of rates. 

Our calculation is more extensive as compared to the earlier work of OHMTS. The overall comparison of rates with OHMTS is fair  for all nuclei at low temperatures where experimental transitions were available. However at higher temperatures, differences begin to show up and go up to many orders of magnitude. Also for cases of negative Q-values the differences get prominent. We attribute these differences to the fact that, firstly, we considered the particle emission processes from excited states which constrained our excitation energies in the parent nuclei,  and secondly, the two different microscopic theories yield different values for the nuclear matrix elements and excitation energies. In the light of these differences we present our rates at precisely the same temperature and density grid points as OHMTS so that a direct comparison be easily available where possible. In the next phase of our program we plan to present the weak rates for $\mathit{fp}$- shell nuclei in stellar matter.\vspace{0.2in}

\textit{\large Acknowledgment}\\
The authors thank Dr. M. Hirsch with whom they had many fruitful discussions right from the beginning of this work. Dr. M. Hirsch also suggested many interesting solutions to the problems encountered during this work and was also kind enough to read the manuscript.

\clearpage
\onecolumn
\oddsidemargin  -0.4in
\begin{center}
\Large{\hspace*{1.0cm}TABLE A}
\end{center}
\vspace*{0.5cm}
\footnotesize
\begin{tabular}{|c||l|l|l|l|l|l|l|l|l|l|l|l|l|} \hline
A/Z & 8 & 9 & 10 & 11 & 12 & 13 & 14 & 15 & 16 & 17 & 18 & 19 & 20\\ \hline\hline
16  &   &   &    &    &    &    &    &    &    &    &    &    &   \\
17  & O$\rightharpoonup$ & $\leftharpoonup$F & & & & & & & & & & & \\
18  & O$\rightharpoonup$ & $\leftharpoonup$F$\rightharpoonup$ & $\Leftarrow$Ne$\rightarrow$ & & & & & & & & & & \\ 
19 & O$\rightharpoonup$ & $\leftharpoonup$F$\rightharpoonup$ & $\Leftarrow$Ne$\Rightarrow$ & $\Leftarrow$Na & & & & & & & & &\\
20 & O$\rightharpoonup$ & $\leftharpoonup$F$\rightharpoonup$ & $\Leftarrow$Ne$\Rightarrow$ & $\Leftarrow$Na$\Rightarrow$ & $\Leftarrow$Mg & & & & & & & &\\ 
21 & O$\rightharpoonup$ & $\leftharpoonup$F$\rightharpoonup$ & $\Leftarrow$Ne$\Rightarrow$ & $\Leftarrow$Na$\Rightarrow$ & $\Leftarrow$Mg$\rightarrow$ & & & & & & & &\\
22 & & \hspace*{0.35cm}F$\rightharpoonup$ & $\Leftarrow$Ne$\Rightarrow$ & $\Leftarrow$Na$\Rightarrow$ & $\Leftarrow$Mg$\rightarrow$ & & & & & & & &\\ 
23 & & \hspace*{0.35cm}F$\rightharpoonup$ & $\Leftarrow$Ne$\Rightarrow$ & $\Leftarrow$Na$\Rightarrow$ & $\Leftarrow$Mg$\Rightarrow$ & $\Leftarrow$Al$\rightarrow$ & & & & & & & \\ 
24 & & &  $\leftarrow$Ne$\Rightarrow$ & $\Leftarrow$Na$\Rightarrow$ & $\Leftarrow$Mg$\Rightarrow$ & $\Leftarrow$Al$\Rightarrow$ & $\Leftarrow$Si$\rightarrow$ & & & & & &\\
25  & & &  $\leftarrow$Ne$\Rightarrow$ & $\Leftarrow$Na$\Rightarrow$ & $\Leftarrow$Mg$\Rightarrow$ & $\Leftarrow$Al$\Rightarrow$ & $\Leftarrow$Si$\rightarrow$ & & & & & &\\
26 & & & & $\leftarrow$Na$\Rightarrow$ & $\Leftarrow$Mg$\Rightarrow$ & $\Leftarrow$Al$\Rightarrow$ & $\Leftarrow$Si$\rightarrow$ & & & & & & \\ 
27 & & & & $\leftarrow$Na$\Rightarrow$ & $\Leftarrow$Mg$\Rightarrow$ & $\Leftarrow$Al$\Rightarrow$ & $\Leftarrow$Si$\Rightarrow$ & $\Leftarrow$P$\rightarrow$ & & & & &\\
28 & & & & $\leftarrow$Na$\Rightarrow$ & $\Leftarrow$Mg$\Rightarrow$ & $\Leftarrow$Al$\Rightarrow$ & $\Leftarrow$Si$\Rightarrow$ & $\Leftarrow$P$\Rightarrow$ & $\Leftarrow$S$\rightarrow$ & & & & \\ 
29 & & & & $\leftarrow$Na$\Rightarrow$ & $\Leftarrow$Mg$\Rightarrow$ & $\Leftarrow$Al$\Rightarrow$ & $\Leftarrow$Si$\Rightarrow$ & $\Leftarrow$P$\Rightarrow$ & $\Leftarrow$S$\rightarrow$ & & & & \\ 
30 & & & & & & $\leftarrow$Al$\Rightarrow$ & $\Leftarrow$Si$\Rightarrow$ & $\Leftarrow$P$\Rightarrow$ & $\Leftarrow$S$\rightarrow$ & & & & \\ 
31 & & & & & & $\leftarrow$Al$\Rightarrow$ & $\Leftarrow$Si$\Rightarrow$ & $\Leftarrow$P$\Rightarrow$ & $\Leftarrow$S$\Rightarrow$ & $\Leftarrow$Cl$\rightarrow$ & & & \\
32 & & & & & & &$\leftarrow$Si$\Rightarrow$ & $\Leftarrow$P$\Rightarrow$ & $\Leftarrow$S$\Rightarrow$ & $\Leftarrow$Cl$\Rightarrow$ & $\Leftarrow$Ar$\rightarrow$ & & \\
33 & & & & & & &$\leftarrow$Si$\Rightarrow$ & $\Leftarrow$P$\Rightarrow$ & $\Leftarrow$S$\Rightarrow$ & $\Leftarrow$Cl$\Rightarrow$ & $\Leftarrow$Ar$\rightarrow$ & & \\
34 & & & & & & &$\leftarrow$Si$\Rightarrow$ & $\Leftarrow$P$\Rightarrow$ & $\Leftarrow$S$\Rightarrow$ & $\Leftarrow$Cl$\Rightarrow$ & $\Leftarrow$Ar$\rightarrow$ & & \\
35 & & & & & & & & $\leftarrow$P$\Rightarrow$ & $\Leftarrow$S$\Rightarrow$ & $\Leftarrow$Cl$\Rightarrow$ & $\Leftarrow$Ar$\Rightarrow$ & $\Leftarrow$K$\rightarrow$ & \\
36 & & & & & & & & & $\leftarrow$S$\Rightarrow$ & $\Leftarrow$Cl$\Rightarrow$ & $\Leftarrow$Ar$\Rightarrow$ & $\Leftarrow$K$\Rightarrow$ & $\Leftarrow$Ca$\rightarrow$\\ 
37 & & & & & & & & & & $\leftarrow$Cl$\Rightarrow$ & $\Leftarrow$Ar$\Rightarrow$ & $\Leftarrow$K$\Rightarrow$ & $\Leftarrow$Ca$\rightarrow$\\
38 & & & & & & & & & & & $\leftarrow$Ar$\Rightarrow$ & $\Leftarrow$K$\Rightarrow$ & $\Leftarrow$Ca$\rightarrow$\\
39 & & & & & & & & & & & & $\leftarrow$K$\Rightarrow$ & $\Leftarrow$Ca$\rightarrow$\\ 
40 & & & & & & & & & & & & & \\ \hline          
\end{tabular} 
\vspace*{0.5in}\\
$\Rightarrow$ \hspace*{0.35cm} $\beta^{-}$ decay, here and in OHMTS\\
$\rightarrow$ \hspace*{0.35cm} $\beta^{-}$ decay, here only\\
$\rightharpoonup$ \hspace*{0.35cm} $\beta^{-}$ decay, in OHMTS only\\
$\Leftarrow$ \hspace*{0.35cm} $\beta^{+}$ decay, here and in OHMTS\\
$\leftarrow$ \hspace*{0.35cm} $\beta^{+}$ decay, here only\\
$\leftharpoonup$ \hspace*{0.35cm} $\beta^{+}$ decay, in OHMTS only\\

\clearpage
\normalsize
\begin{center}
\textbf{\Large EXPLANATION OF TABLE}
\end{center}
\vspace{0.7in}
\textbf{\large TABLE. $\mathbf{sd}$- Shell Nuclei Weak Rates in Stellar Environment}
\vspace{0.4in}
\noindent
\newline
The calculated weak interaction rates [Eqs. (15) and (18)-(20)] are all tabulated in log$_{10} \lambda$. The probabilities of $\beta$-delayed proton (neutron) emission [Eq. ~(21)] are also tabulated in logarithmic scale. It should be noted that the probabilities are calculated only to one significant figure and are given up to three places of decimal only for designing purposes. All rates listed for a particular direction concern the parent nucleus except for the last two columns which concern the daughter nucleus. For each daughter nucleus, either the proton energy rate and probability of $\beta$-delayed proton emission is stated (if $S_{p} < S_{n}$) or the neutron energy rate and probability of $\beta$-delayed neutron emission is stated (if $S_{n} < S_{p}$). In the table -100 means that the rate (or the probability) is smaller than $10^{-100}$. \\
\begin {tabbing}
\textsf{Q} \hspace{0.7in}        \= Mass of parent minus mass of daughter nucleus\\
S$_{\mathsf{nP}}$                       \> Separation energy of neutron for parent nucleus\\
S$_{\mathsf{pP}}$                       \> Separation energy of proton for parent nucleus\\
S$_{\mathsf{nD}}$                       \> Separation energy of neutron for daughter nucleus\\
S$_{\mathsf{pD}}$                     \> Separation energy of neutron for daughter nucleus\\
\textsf{ADen}                    \> log $(\rho Y_{e})$ (g.cm$^{-3}$), where $\rho$ is the density of the baryon, and $Y_{e}$ is the ratio of the\\
                        \> electron number to the baryon number\\
\textsf{T9}                      \> Temperature in units of $10^{9}$ K\\
\textsf{EFermi}                  \> Total Fermi energy of electron and positron, including the rest mass (MeV)\\
\textsf{E+Cap}                   \> Positron capture rate (s$^{-1}$)\\
\textsf{E-Dec}                   \> Electron decay rate (s$^{-1}$)\\ 
\textsf{ANuEn}                   \> Anti-Neutrino energy loss rate (MeV.s$^{-1}$)\\
\textsf{GamEn}                 \> Gamma ray heating rate (MeV.s$^{-1}$), tabulated separately for $\beta^{+}$ direction \\
                        \> and $\beta^{-}$ direction \\
\textsf{E-Cap}                   \> Electron capture rate (s$^{-1}$)\\
\textsf{E+Dec}                   \> Positron decay rate (s$^{-1}$)\\
\textsf{NuEn}                 \> Neutrino energy loss rate (MeV.s$^{-1}$)\\     
\textsf{ProEn}                  \> Energy rate of $\beta$-delayed proton (MeV.s$^{-1}$)\\
\textsf{NeuEn}                  \> Energy rate of $\beta$-delayed neutron (MeV.s$^{-1}$)\\
\textsf{PPEm}                  \> Probability of $\beta$-delayed proton emission\\
\textsf{PNEm}                  \> Probability of $\beta$-delayed neutron emission\\             
\end{tabbing}
\sffamily \normalsize
\noindent
\begin{tabular}{rr|rrrrrr}
\multicolumn{7}{c}{\normalsize{\hspace*{3.5cm}$^{  18}_{10}$ NE $_{  8} \rightarrow \hspace*{0.2cm}^{  18}_{11}$ NA $_{  7}$}}\\
\hline
Aden& T9&  E+Cap &   E-Dec &  ANuEn &  GamEn &  ProEn &  PPEm \\
\hline
     3.0&  1.0 &    -100 &    -100 &    -100 &    -100 &    -100 &-100\\
     3.0&  3.0 & -35.671 &    -100 & -35.521 &    -100 & -35.950 &  0.000 \\
     3.0& 30.0 &  -1.537 &    -100 &  -0.571 &    -100 &  -1.597 &  0.000 \\
     7.0&  1.0 &    -100 &    -100 &    -100 &    -100 &    -100 &-100\\
     7.0&  3.0 & -37.387 &    -100 & -37.237 &    -100 & -37.665 &  0.000 \\
     7.0& 30.0 &  -1.541 &    -100 &  -0.574 &    -100 &  -1.600 &  0.000 \\
    11.0&  1.0 &    -100 &    -100 &    -100 &    -100 &    -100 &-100\\
    11.0&  3.0 & -75.864 &    -100 & -75.714 &    -100 & -76.142 &  0.000 \\
    11.0& 30.0 &  -5.404 &    -100 &  -4.437 &    -100 &  -5.463 &  0.000 \\
\hline
\end{tabular}\\
\vspace{0.20cm}\\
\begin{tabular}{rr|rrrrrr}
\multicolumn{7}{c}{\normalsize{\hspace*{3.5cm}$^{  18}_{10}$ NE $_{  8} \rightarrow \hspace*{0.2cm}^{  18}_{ 9}$ F  $_{  9}$}}\\
\hline
Aden& T9&  E-Cap &   E+Dec &  NuEn &  GamEn &  ProEn &  PPEm \\
\hline
     3.0&  1.0 &  -4.323 &  -0.387 &  -0.122 &  -1.466 & -17.237 &-20.000 \\
     3.0&  3.0 &  -1.918 &  -0.389 &  -0.093 &  -1.434 &  -7.351 & -7.000 \\
     3.0& 30.0 &   2.223 &  -0.553 &   3.391 &   1.876 &   1.675 & -0.900 \\
     7.0&  1.0 &  -0.511 &  -0.387 &   0.345 &  -0.961 & -11.417 &-10.000 \\
     7.0&  3.0 &  -0.470 &  -0.387 &   0.390 &  -0.923 &  -5.636 & -5.000 \\
     7.0& 30.0 &   2.226 &  -0.552 &   3.395 &   1.880 &   1.678 & -0.900 \\
    11.0&  1.0 &   4.917 &  -0.387 &   6.252 &   4.605 &   4.629 & -0.700 \\
    11.0&  3.0 &   4.918 &  -0.387 &   6.253 &   4.606 &   4.630 & -0.700 \\
    11.0& 30.0 &   4.984 &  -0.387 &   6.369 &   4.673 &   4.727 & -0.700 \\
\hline
\end{tabular}\\
\vspace{0.20cm}\\
\noindent
\begin{tabular}{rr|rrrrrr}
\multicolumn{7}{c}{\normalsize{\hspace*{3.5cm}$^{  19}_{10}$ NE $_{  9} \rightarrow \hspace*{0.2cm}^{  19}_{11}$ NA $_{  8}$}}\\
\hline
Aden& T9&  E+Cap &   E-Dec &  ANuEn &  GamEn &  ProEn &  PPEm \\
\hline
     3.0&  1.0 & -60.842 &    -100 & -60.791 & -70.845 & -63.402 & -2.000 \\
     3.0&  3.0 & -20.619 &    -100 & -20.465 & -30.705 & -21.693 & -0.700 \\
     3.0& 30.0 &  -0.193 &    -100 &   0.813 & -10.947 &   0.315 & -0.080 \\
     7.0&  1.0 & -66.662 &    -100 & -66.611 & -76.665 & -69.222 & -2.000 \\
     7.0&  3.0 & -22.335 &    -100 & -22.180 & -32.421 & -23.408 & -0.700 \\
     7.0& 30.0 &  -0.197 &    -100 &   0.810 & -10.951 &   0.311 & -0.080 \\
    11.0&  1.0 &    -100 &    -100 &    -100 &    -100 &    -100 &-100\\
    11.0&  3.0 & -60.812 &    -100 & -60.658 & -70.898 & -61.886 & -0.700 \\
    11.0& 30.0 &  -4.059 &    -100 &  -3.053 & -14.813 &  -3.552 & -0.080 \\
\hline
\end{tabular}\\
\vspace{0.20cm}\\
\begin{tabular}{rr|rrrrrr}
\multicolumn{7}{c}{\normalsize{\hspace*{3.5cm}$^{  19}_{10}$ NE $_{  9} \rightarrow \hspace*{0.2cm}^{  19}_{ 9}$ F  $_{ 10}$}}\\
\hline
Aden& T9&  E-Cap &   E+Dec &  NuEn &  GamEn &  ProEn &  PPEm \\
\hline
     3.0&  1.0 &  -4.832 &  -1.403 &  -1.305 &  -5.709 & -33.604 &-30.000 \\
     3.0&  3.0 &  -2.407 &  -1.409 &  -1.196 &  -4.075 & -12.256 &-10.000 \\
     3.0& 30.0 &   1.866 &  -1.300 &   2.973 &   2.175 &   1.890 & -0.500 \\
     7.0&  1.0 &  -1.000 &  -1.403 &  -0.372 &  -2.857 & -27.790 &-30.000 \\
     7.0&  3.0 &  -0.949 &  -1.405 &  -0.296 &  -2.486 & -10.543 &-10.000 \\
     7.0& 30.0 &   1.870 &  -1.299 &   2.977 &   2.179 &   1.893 & -0.500 \\
    11.0&  1.0 &   4.861 &  -1.403 &   6.126 &   5.374 &   4.309 & -1.000 \\
    11.0&  3.0 &   4.860 &  -1.405 &   6.125 &   5.373 &   4.316 & -1.000 \\
    11.0& 30.0 &   4.811 &  -1.170 &   6.121 &   5.147 &   5.065 & -0.300 \\
\hline
\end{tabular}\\
\vspace{0.20cm}\\
\noindent
\begin{tabular}{rr|rrrrrr}
\multicolumn{7}{c}{\normalsize{\hspace*{3.5cm}$^{  19}_{11}$ NA $_{  8} \rightarrow \hspace*{0.2cm}^{  19}_{10}$ NE $_{  9}$}}\\
\hline
Aden& T9&  E-Cap &   E+Dec &  NuEn &  GamEn &  ProEn &  PPEm \\
\hline
     3.0&  1.0 &  -4.335 &   0.834 &   1.509 &   0.877 &  -1.986 & -3.000 \\
     3.0&  3.0 &  -1.934 &   0.833 &   1.507 &   0.909 &  -1.492 & -2.000 \\
     3.0& 30.0 &   2.307 &   1.086 &   3.514 &   2.387 &   2.861 & -0.200 \\
     7.0&  1.0 &  -0.562 &   0.834 &   1.543 &   0.899 &  -1.230 & -2.000 \\
     7.0&  3.0 &  -0.502 &   0.833 &   1.544 &   0.933 &  -0.966 & -2.000 \\
     7.0& 30.0 &   2.310 &   1.087 &   3.518 &   2.390 &   2.864 & -0.200 \\
    11.0&  1.0 &   4.773 &   0.834 &   6.092 &   4.552 &   5.432 & -0.100 \\
    11.0&  3.0 &   4.809 &   0.833 &   6.132 &   4.636 &   5.449 & -0.100 \\
    11.0& 30.0 &   5.071 &   1.145 &   6.456 &   4.942 &   5.777 & -0.100 \\
\hline
\end{tabular}\\
\vspace{0.20cm}\\
\noindent
\begin{tabular}{rr|rrrrrr}
\multicolumn{7}{c}{\normalsize{\hspace*{3.5cm}$^{  20}_{10}$ NE $_{ 10} \rightarrow \hspace*{0.2cm}^{  20}_{11}$ NA $_{  9}$}}\\
\hline
Aden& T9&  E+Cap &   E-Dec &  ANuEn &  GamEn &  ProEn &  PPEm \\
\hline
     3.0&  1.0 & -74.144 &    -100 & -74.092 & -76.615 & -88.667 &-10.000 \\
     3.0&  3.0 & -24.788 &    -100 & -24.627 & -25.746 & -29.461 & -5.000 \\
     3.0& 30.0 &   0.330 &    -100 &   1.363 &  -0.120 &   0.590 & -0.300 \\
     7.0&  1.0 & -79.964 &    -100 & -79.912 & -82.435 & -94.487 &-10.000 \\
     7.0&  3.0 & -26.504 &    -100 & -26.342 & -27.461 & -31.176 & -5.000 \\
     7.0& 30.0 &   0.327 &    -100 &   1.359 &  -0.123 &   0.586 & -0.300 \\
    11.0&  1.0 &    -100 &    -100 &    -100 &    -100 &    -100 &-100\\
    11.0&  3.0 & -64.981 &    -100 & -64.820 & -65.939 & -69.654 & -5.000 \\
    11.0& 30.0 &  -3.534 &    -100 &  -2.502 &  -3.983 &  -3.276 & -0.300 \\
\hline
\end{tabular}\\
\vspace{0.20cm}\\
\begin{tabular}{rr|rrrrrr}
\multicolumn{7}{c}{\normalsize{\hspace*{3.5cm}$^{  20}_{10}$ NE $_{ 10} \rightarrow \hspace*{0.2cm}^{  20}_{ 9}$ F  $_{ 11}$}}\\
\hline
Aden & T9& E-Cap &   E+Dec &  NuEn &  GamEn &  NeuEn &  PNEm \\
\hline
     3.0&  1.0 & -42.254 & -51.342 & -42.831 & -44.181 & -75.784 &-30.000 \\
     3.0&  3.0 & -15.562 & -18.407 & -15.632 & -16.572 & -26.133 &-10.000 \\
     3.0& 30.0 &   1.144 &  -3.325 &   2.226 &   1.487 &   0.979 & -0.700 \\
     7.0&  1.0 & -36.434 & -51.342 & -37.011 & -38.370 & -69.977 &-30.000 \\
     7.0&  3.0 & -13.851 & -18.400 & -13.920 & -14.886 & -24.419 &-10.000 \\
     7.0& 30.0 &   1.148 &  -3.323 &   2.230 &   1.490 &   0.983 & -0.700 \\
    11.0&  1.0 &   3.338 & -51.342 &   4.358 &   3.801 &   2.269 & -1.000 \\
    11.0&  3.0 &   3.339 & -18.399 &   4.361 &   3.802 &   2.276 & -1.000 \\
    11.0& 30.0 &   4.163 &  -3.132 &   5.441 &   4.465 &   4.290 & -0.500 \\
\hline
\end{tabular}\\
\vspace{0.20cm}\\
\noindent
\begin{tabular}{rr|rrrrrr}
\multicolumn{7}{c}{\normalsize{\hspace*{3.5cm}$^{  20}_{11}$ NA $_{  9} \rightarrow \hspace*{0.2cm}^{  20}_{12}$ MG $_{  8}$}}\\
\hline
Aden& T9&  E+Cap &   E-Dec &  ANuEn &  GamEn &  ProEn &  PPEm \\
\hline
     3.0&  1.0 & -86.117 &    -100 & -86.067 &    -100 & -85.655 &  0.000 \\
     3.0&  3.0 & -28.863 &    -100 & -28.712 &    -100 & -28.383 &  0.000 \\
     3.0& 30.0 &  -0.835 &    -100 &   0.141 &    -100 &  -0.098 &  0.000 \\
     7.0&  1.0 & -91.937 &    -100 & -91.887 &    -100 & -91.475 &  0.000 \\
     7.0&  3.0 & -30.579 &    -100 & -30.428 &    -100 & -30.098 &  0.000 \\
     7.0& 30.0 &  -0.839 &    -100 &   0.138 &    -100 &  -0.101 &  0.000 \\
    11.0&  1.0 &    -100 &    -100 &    -100 &    -100 &    -100 &-100\\
    11.0&  3.0 & -69.056 &    -100 & -68.905 &    -100 & -68.576 &  0.000 \\
    11.0& 30.0 &  -4.702 &    -100 &  -3.725 &    -100 &  -3.964 &  0.000 \\
\hline
\end{tabular}\\
\vspace{0.20cm}\\
\begin{tabular}{rr|rrrrrr}
\multicolumn{7}{c}{\normalsize{\hspace*{3.5cm}$^{  20}_{11}$ NA $_{  9} \rightarrow \hspace*{0.2cm}^{  20}_{10}$ NE $_{ 10}$}}\\
\hline
Aden& T9&  E-Cap &   E+Dec &  NuEn &  GamEn &  ProEn &  PPEm \\
\hline
     3.0&  1.0 &  -4.398 &   0.408 &   1.066 &   1.030 &  -8.025 & -8.000 \\
     3.0&  3.0 &  -2.070 &   0.320 &   0.959 &   0.996 &  -4.589 & -5.000 \\
     3.0& 30.0 &   2.125 &   0.330 &   3.295 &   2.907 &   2.301 & -0.400 \\
     7.0&  1.0 &  -0.602 &   0.408 &   1.123 &   1.104 &  -4.032 & -4.000 \\
     7.0&  3.0 &  -0.628 &   0.320 &   1.027 &   1.069 &  -3.069 & -3.000 \\
     7.0& 30.0 &   2.129 &   0.330 &   3.299 &   2.910 &   2.304 & -0.400 \\
    11.0&  1.0 &   4.801 &   0.408 &   6.129 &   5.627 &   4.945 & -0.500 \\
    11.0&  3.0 &   4.790 &   0.320 &   6.109 &   5.572 &   5.023 & -0.400 \\
    11.0& 30.0 &   4.927 &   0.421 &   6.293 &   5.592 &   5.337 & -0.300 \\
\hline
\end{tabular}\\
\vspace{0.20cm}\\
\noindent
\begin{tabular}{rr|rrrrrr}
\multicolumn{7}{c}{\normalsize{\hspace*{3.5cm}$^{  20}_{12}$ MG $_{  8} \rightarrow \hspace*{0.2cm}^{  20}_{11}$ NA $_{  9}$}}\\
\hline
Aden& T9&  E-Cap &   E+Dec &  NuEn &  GamEn &  ProEn &  PPEm \\
\hline
     3.0&  1.0 &  -3.629 &   1.193 &   1.756 &    -100 &   0.977 &  0.000 \\
     3.0&  3.0 &  -1.264 &   1.193 &   1.759 &    -100 &   0.979 &  0.000 \\
     3.0& 30.0 &   2.657 &   1.095 &   3.876 &    -100 &   3.148 &  0.000 \\
     7.0&  1.0 &   0.145 &   1.193 &   1.833 &    -100 &   1.044 &  0.000 \\
     7.0&  3.0 &   0.170 &   1.193 &   1.839 &    -100 &   1.052 &  0.000 \\
     7.0& 30.0 &   2.661 &   1.096 &   3.880 &    -100 &   3.151 &  0.000 \\
    11.0&  1.0 &   5.291 &   1.193 &   6.665 &    -100 &   5.891 &  0.000 \\
    11.0&  3.0 &   5.291 &   1.193 &   6.666 &    -100 &   5.892 &  0.000 \\
    11.0& 30.0 &   5.349 &   1.193 &   6.767 &    -100 &   5.964 &  0.000 \\
\hline
\end{tabular}\\
\vspace{0.20cm}\\
\noindent
\begin{tabular}{rr|rrrrrr}
\multicolumn{7}{c}{\normalsize{\hspace*{3.5cm}$^{  21}_{10}$ NE $_{ 11} \rightarrow \hspace*{0.2cm}^{  21}_{11}$ NA $_{ 10}$}}\\
\hline
Aden& T9&  E+Cap &   E-Dec &  ANuEn &  GamEn &  ProEn &  PPEm \\
\hline
     3.0&  1.0 & -22.544 & -30.607 & -22.612 & -26.151 & -39.330 &-20.000 \\
     3.0&  3.0 &  -8.379 & -12.582 &  -8.213 & -10.482 & -13.493 & -5.000 \\
     3.0& 30.0 &   0.760 &  -5.147 &   1.814 &  -0.650 &   1.213 & -0.200 \\
     7.0&  1.0 & -28.364 & -30.966 & -28.431 & -31.375 & -45.150 &-20.000 \\
     7.0&  3.0 & -10.094 & -12.804 &  -9.928 & -12.163 & -15.208 & -5.000 \\
     7.0& 30.0 &   0.756 &  -5.148 &   1.811 &  -0.653 &   1.210 & -0.200 \\
    11.0&  1.0 &    -100 &    -100 &    -100 &    -100 &    -100 &-100\\
    11.0&  3.0 & -48.571 & -50.488 & -48.404 & -50.511 & -53.686 & -5.000 \\
    11.0& 30.0 &  -3.102 &  -8.601 &  -2.050 &  -4.504 &  -2.652 & -0.200 \\
\hline
\end{tabular}\\
\vspace{0.20cm}\\
\begin{tabular}{rr|rrrrrr}
\multicolumn{7}{c}{\normalsize{\hspace*{3.5cm}$^{  21}_{10}$ NE $_{ 11} \rightarrow \hspace*{0.2cm}^{  21}_{ 9}$ F  $_{ 12}$}}\\
\hline
Aden & T9& E-Cap &   E+Dec &  NuEn &  GamEn &  NeuEn &  PNEm \\
\hline
     3.0&  1.0 & -34.755 &    -100 & -35.321 & -36.880 & -76.540 &-40.000 \\
     3.0&  3.0 & -12.543 &    -100 & -12.606 & -13.278 & -26.419 &-10.000 \\
     3.0& 30.0 &   0.744 &    -100 &   1.776 &   0.965 &   0.150 & -1.000 \\
     7.0&  1.0 & -28.935 &    -100 & -29.502 & -31.061 & -70.720 &-40.000 \\
     7.0&  3.0 & -10.827 &    -100 & -10.891 & -11.563 & -24.704 &-10.000 \\
     7.0& 30.0 &   0.748 &    -100 &   1.780 &   0.969 &   0.154 & -1.000 \\
    11.0&  1.0 &   3.814 &    -100 &   4.831 &   4.392 &   2.874 & -1.000 \\
    11.0&  3.0 &   3.809 &    -100 &   4.844 &   4.343 &   2.940 & -1.000 \\
    11.0& 30.0 &   3.970 &    -100 &   5.177 &   4.305 &   3.714 & -0.700 \\
\hline
\end{tabular}\\
\vspace{0.20cm}\\
\noindent
\begin{tabular}{rr|rrrrrr}
\multicolumn{7}{c}{\normalsize{\hspace*{3.5cm}$^{  21}_{11}$ NA $_{ 10} \rightarrow \hspace*{0.2cm}^{  21}_{12}$ MG $_{  9}$}}\\
\hline
Aden& T9&  E+Cap &   E-Dec &  ANuEn &  GamEn &  ProEn &  PPEm \\
\hline
     3.0&  1.0 & -69.955 &    -100 & -69.904 & -73.360 & -89.634 &-20.000 \\
     3.0&  3.0 & -23.416 &    -100 & -23.263 & -24.556 & -31.375 & -7.000 \\
     3.0& 30.0 &  -0.286 &    -100 &   0.707 &  -0.321 &  -0.404 & -0.900 \\
     7.0&  1.0 & -75.775 &    -100 & -75.724 & -79.180 & -95.454 &-20.000 \\
     7.0&  3.0 & -25.132 &    -100 & -24.979 & -26.271 & -33.091 & -7.000 \\
     7.0& 30.0 &  -0.289 &    -100 &   0.704 &  -0.324 &  -0.407 & -0.900 \\
    11.0&  1.0 &    -100 &    -100 &    -100 &    -100 &    -100 &-100\\
    11.0&  3.0 & -63.609 &    -100 & -63.456 & -64.748 & -71.568 & -7.000 \\
    11.0& 30.0 &  -4.152 &    -100 &  -3.159 &  -4.187 &  -4.270 & -0.900 \\
\hline
\end{tabular}\\
\vspace{0.20cm}\\
\begin{tabular}{rr|rrrrrr}
\multicolumn{7}{c}{\normalsize{\hspace*{3.5cm}$^{  21}_{11}$ NA $_{ 10} \rightarrow \hspace*{0.2cm}^{  21}_{10}$ NE $_{ 11}$}}\\
\hline
Aden & T9& E-Cap &   E+Dec &  NuEn &  GamEn &  NeuEn &  PNEm \\
\hline
     3.0&  1.0 &  -5.077 &  -1.518 &  -1.370 &  -3.177 & -24.764 &-20.000 \\
     3.0&  3.0 &  -2.715 &  -1.584 &  -1.354 &  -2.687 &  -9.529 & -8.000 \\
     3.0& 30.0 &   1.510 &  -2.013 &   2.621 &   1.376 &   1.517 & -0.400 \\
     7.0&  1.0 &  -1.257 &  -1.518 &  -0.577 &  -2.398 & -18.944 &-20.000 \\
     7.0&  3.0 &  -1.260 &  -1.580 &  -0.567 &  -1.936 &  -7.814 & -6.000 \\
     7.0& 30.0 &   1.513 &  -2.011 &   2.624 &   1.380 &   1.520 & -0.400 \\
    11.0&  1.0 &   4.496 &  -1.518 &   5.755 &   3.602 &   4.621 & -0.300 \\
    11.0&  3.0 &   4.501 &  -1.580 &   5.758 &   4.092 &   4.628 & -0.300 \\
    11.0& 30.0 &   4.451 &  -1.832 &   5.759 &   4.201 &   4.732 & -0.200 \\
\hline
\end{tabular}\\
\vspace{0.20cm}\\
\noindent
\begin{tabular}{rr|rrrrrr}
\multicolumn{7}{c}{\normalsize{\hspace*{3.5cm}$^{  21}_{12}$ MG $_{  9} \rightarrow \hspace*{0.2cm}^{  21}_{13}$ AL $_{  8}$}}\\
\hline
Aden& T9&  E+Cap &   E-Dec &  ANuEn &  GamEn &  ProEn &  PPEm \\
\hline
     3.0&  1.0 & -81.148 &    -100 & -81.097 & -91.148 & -86.165 & -5.000 \\
     3.0&  3.0 & -27.494 &    -100 & -27.342 & -37.521 & -28.641 & -1.000 \\
     3.0& 30.0 &  -0.844 &    -100 &   0.142 & -11.620 &  -0.532 & -0.080 \\
     7.0&  1.0 & -86.968 &    -100 & -86.917 & -96.968 & -91.985 & -5.000 \\
     7.0&  3.0 & -29.210 &    -100 & -29.057 & -39.237 & -30.356 & -1.000 \\
     7.0& 30.0 &  -0.847 &    -100 &   0.138 & -11.624 &  -0.535 & -0.080 \\
    11.0&  1.0 &    -100 &    -100 &    -100 &    -100 &    -100 &-100\\
    11.0&  3.0 & -67.687 &    -100 & -67.535 & -77.714 & -68.834 & -1.000 \\
    11.0& 30.0 &  -4.710 &    -100 &  -3.725 & -15.486 &  -4.398 & -0.080 \\
\hline
\end{tabular}\\
\vspace{0.20cm}\\
\begin{tabular}{rr|rrrrrr}
\multicolumn{7}{c}{\normalsize{\hspace*{3.5cm}$^{  21}_{12}$ MG $_{  9} \rightarrow \hspace*{0.2cm}^{  21}_{11}$ NA $_{ 10}$}}\\
\hline
Aden& T9&  E-Cap &   E+Dec &  NuEn &  GamEn &  ProEn &  PPEm \\
\hline
     3.0&  1.0 &  -3.881 &   0.858 &   1.535 &   0.276 &   1.072 & -0.300 \\
     3.0&  3.0 &  -1.533 &   0.874 &   1.565 &   0.529 &   1.028 & -0.400 \\
     3.0& 30.0 &   2.531 &   1.061 &   3.723 &   1.468 &   3.432 & -0.030 \\
     7.0&  1.0 &  -0.095 &   0.858 &   1.596 &   0.284 &   1.196 & -0.300 \\
     7.0&  3.0 &  -0.094 &   0.874 &   1.624 &   0.538 &   1.160 & -0.400 \\
     7.0& 30.0 &   2.535 &   1.061 &   3.727 &   1.471 &   3.435 & -0.030 \\
    11.0&  1.0 &   5.225 &   0.858 &   6.570 &   3.283 &   6.119 & -0.008 \\
    11.0&  3.0 &   5.207 &   0.874 &   6.550 &   3.557 &   6.111 & -0.010 \\
    11.0& 30.0 &   5.286 &   1.126 &   6.673 &   3.904 &   6.256 & -0.010 \\
\hline
\end{tabular}\\
\vspace{0.20cm}\\
\noindent
\begin{tabular}{rr|rrrrrr}
\multicolumn{7}{c}{\normalsize{\hspace*{3.5cm}$^{  22}_{10}$ NE $_{ 12} \rightarrow \hspace*{0.2cm}^{  22}_{11}$ NA $_{ 11}$}}\\
\hline
Aden& T9&  E+Cap &   E-Dec &  ANuEn &  GamEn &  ProEn &  PPEm \\
\hline
     3.0&  1.0 & -23.629 & -32.322 & -23.575 & -26.709 & -55.008 &-30.000 \\
     3.0&  3.0 & -10.191 & -10.968 &  -9.942 & -10.083 & -20.083 & -8.000 \\
     3.0& 30.0 &   1.334 &  -0.221 &   2.498 &   1.608 &   1.172 & -0.800 \\
     7.0&  1.0 & -29.449 & -32.409 & -29.394 & -32.445 & -59.314 &-30.000 \\
     7.0&  3.0 & -11.906 & -11.033 & -10.742 & -11.488 & -21.756 & -9.000 \\
     7.0& 30.0 &   1.331 &  -0.222 &   2.495 &   1.605 &   1.169 & -0.800 \\
    11.0&  1.0 &    -100 &    -100 &    -100 &    -100 &    -100 &-100\\
    11.0&  3.0 & -50.384 & -44.938 & -45.084 & -47.663 & -60.159 &-10.000 \\
    11.0& 30.0 &  -2.517 &  -3.302 &  -1.346 &  -2.244 &  -2.690 & -0.900 \\
\hline
\end{tabular}\\
\vspace{0.20cm}\\
\begin{tabular}{rr|rrrrrr}
\multicolumn{7}{c}{\normalsize{\hspace*{3.5cm}$^{  22}_{10}$ NE $_{ 12} \rightarrow \hspace*{0.2cm}^{  22}_{ 9}$ F  $_{ 13}$}}\\
\hline
Aden & T9& E-Cap &   E+Dec &  NuEn &  GamEn &  NeuEn &  PNEm \\
\hline
     3.0&  1.0 & -60.331 &    -100 & -60.897 & -65.075 & -87.805 &-30.000 \\
     3.0&  3.0 & -20.782 &    -100 & -20.835 & -22.014 & -29.430 & -8.000 \\
     3.0& 30.0 &   0.750 &    -100 &   1.791 &   0.869 &   0.419 & -0.600 \\
     7.0&  1.0 & -54.520 &    -100 & -55.087 & -59.255 & -81.985 &-30.000 \\
     7.0&  3.0 & -19.071 &    -100 & -19.122 & -20.299 & -27.714 & -8.000 \\
     7.0& 30.0 &   0.754 &    -100 &   1.795 &   0.873 &   0.423 & -0.600 \\
    11.0&  1.0 &   2.909 &    -100 &   3.793 &   3.242 &   1.314 & -0.900 \\
    11.0&  3.0 &   2.909 &    -100 &   3.796 &   3.242 &   1.317 & -0.900 \\
    11.0& 30.0 &   3.912 &    -100 &   5.142 &   4.026 &   3.873 & -0.400 \\
\hline
\end{tabular}\\
\vspace{0.20cm}\\
\noindent
\begin{tabular}{rr|rrrrrr}
\multicolumn{7}{c}{\normalsize{\hspace*{3.5cm}$^{  22}_{11}$ NA $_{ 11} \rightarrow \hspace*{0.2cm}^{  22}_{12}$ MG $_{ 10}$}}\\
\hline
Aden& T9&  E+Cap &   E-Dec &  ANuEn &  GamEn &  ProEn &  PPEm \\
\hline
     3.0&  1.0 & -28.494 &    -100 & -28.442 &    -100 & -62.574 &-30.000 \\
     3.0&  3.0 &  -9.952 &    -100 &  -9.788 &    -100 & -20.567 &-10.000 \\
     3.0& 30.0 &   0.657 &    -100 &   1.689 &    -100 &   1.064 & -0.300 \\
     7.0&  1.0 & -34.314 &    -100 & -34.262 &    -100 & -68.394 &-30.000 \\
     7.0&  3.0 & -11.668 &    -100 & -11.504 &    -100 & -22.283 &-10.000 \\
     7.0& 30.0 &   0.653 &    -100 &   1.686 &    -100 &   1.061 & -0.300 \\
    11.0&  1.0 &    -100 &    -100 &    -100 &    -100 &    -100 &-100\\
    11.0&  3.0 & -50.145 &    -100 & -49.981 &    -100 & -60.760 &-10.000 \\
    11.0& 30.0 &  -3.207 &    -100 &  -2.176 &    -100 &  -2.802 & -0.300 \\
\hline
\end{tabular}\\
\vspace{0.20cm}\\
\begin{tabular}{rr|rrrrrr}
\multicolumn{7}{c}{\normalsize{\hspace*{3.5cm}$^{  22}_{11}$ NA $_{ 11} \rightarrow \hspace*{0.2cm}^{  22}_{10}$ NE $_{ 12}$}}\\
\hline
Aden & T9& E-Cap &   E+Dec &  NuEn &  GamEn &  NeuEn &  PNEm \\
\hline
     3.0&  1.0 &  -9.579 &  -8.108 &  -8.516 &  -7.990 & -44.311 &-40.000 \\
     3.0&  3.0 &  -7.219 &  -8.302 &  -6.865 &  -7.034 & -15.589 & -8.000 \\
     3.0& 30.0 &   1.618 &  -9.137 &   2.666 &   2.372 &   1.540 & -0.500 \\
     7.0&  1.0 &  -5.674 &  -8.108 &  -5.348 &  -5.569 & -38.491 &-30.000 \\
     7.0&  3.0 &  -5.721 &  -8.275 &  -5.339 &  -5.536 & -13.874 & -7.000 \\
     7.0& 30.0 &   1.621 &  -9.135 &   2.670 &   2.376 &   1.544 & -0.500 \\
    11.0&  1.0 &   4.245 &  -8.108 &   5.393 &   5.062 &   4.151 & -0.800 \\
    11.0&  3.0 &   4.390 &  -8.274 &   5.546 &   5.191 &   4.284 & -0.800 \\
    11.0& 30.0 &   4.711 &  -8.892 &   5.966 &   5.399 &   4.959 & -0.400 \\
\hline
\end{tabular}\\
\vspace{0.20cm}\\
\noindent
\begin{tabular}{rr|rrrrrr}
\multicolumn{7}{c}{\normalsize{\hspace*{3.5cm}$^{  22}_{12}$ MG $_{ 10} \rightarrow \hspace*{0.2cm}^{  22}_{13}$ AL $_{  9}$}}\\
\hline
Aden& T9&  E+Cap &   E-Dec &  ANuEn &  GamEn &  ProEn &  PPEm \\
\hline
     3.0&  1.0 & -97.946 &    -100 & -97.896 & -99.647 &    -100 & -4.000 \\
     3.0&  3.0 & -33.110 &    -100 & -32.960 & -34.888 & -33.899 & -0.800 \\
     3.0& 30.0 &  -1.050 &    -100 &  -0.083 &  -3.815 &  -0.854 & -0.040 \\
     7.0&  1.0 &    -100 &    -100 &    -100 &    -100 &    -100 &-100\\
     7.0&  3.0 & -34.826 &    -100 & -34.675 & -36.603 & -35.614 & -0.800 \\
     7.0& 30.0 &  -1.054 &    -100 &  -0.086 &  -3.819 &  -0.857 & -0.040 \\
    11.0&  1.0 &    -100 &    -100 &    -100 &    -100 &    -100 &-100\\
    11.0&  3.0 & -73.303 &    -100 & -73.153 & -75.081 & -74.092 & -0.800 \\
    11.0& 30.0 &  -4.917 &    -100 &  -3.949 &  -7.682 &  -4.720 & -0.040 \\
\hline
\end{tabular}\\
\vspace{0.20cm}\\
\begin{tabular}{rr|rrrrrr}
\multicolumn{7}{c}{\normalsize{\hspace*{3.5cm}$^{  22}_{12}$ MG $_{ 10} \rightarrow \hspace*{0.2cm}^{  22}_{11}$ NA $_{ 11}$}}\\
\hline
Aden& T9&  E-Cap &   E+Dec &  NuEn &  GamEn &  ProEn &  PPEm \\
\hline
     3.0&  1.0 &  -4.240 &  -0.438 &  -0.172 &  -0.820 & -17.921 &-20.000 \\
     3.0&  3.0 &  -1.832 &  -0.440 &  -0.135 &  -0.782 &  -7.860 & -6.000 \\
     3.0& 30.0 &   2.415 &  -0.604 &   3.568 &   2.614 &   1.615 & -1.000 \\
     7.0&  1.0 &  -0.428 &  -0.438 &   0.357 &  -0.275 & -12.101 &-10.000 \\
     7.0&  3.0 &  -0.379 &  -0.438 &   0.411 &  -0.225 &  -6.145 & -5.000 \\
     7.0& 30.0 &   2.419 &  -0.602 &   3.571 &   2.617 &   1.618 & -1.000 \\
    11.0&  1.0 &   5.124 &  -0.438 &   6.447 &   5.330 &   4.628 & -0.800 \\
    11.0&  3.0 &   5.125 &  -0.438 &   6.448 &   5.330 &   4.629 & -0.800 \\
    11.0& 30.0 &   5.191 &  -0.438 &   6.566 &   5.395 &   4.747 & -0.700 \\
\hline
\end{tabular}\\
\vspace{0.20cm}\\
\noindent
\begin{tabular}{rr|rrrrrr}
\multicolumn{7}{c}{\normalsize{\hspace*{3.5cm}$^{  23}_{10}$ NE $_{ 13} \rightarrow \hspace*{0.2cm}^{  23}_{11}$ NA $_{ 12}$}}\\
\hline
Aden& T9&  E+Cap &   E-Dec &  ANuEn &  GamEn &  ProEn &  PPEm \\
\hline
     3.0&  1.0 &  -6.794 &  -1.742 &  -1.378 &  -2.516 & -29.200 &-30.000 \\
     3.0&  3.0 &  -3.403 &  -1.457 &  -1.070 &  -1.782 & -10.610 & -9.000 \\
     3.0& 30.0 &   1.705 &  -0.675 &   2.827 &   2.356 &   1.381 & -0.700 \\
     7.0&  1.0 & -12.614 &  -1.784 &  -1.446 &  -2.582 & -33.886 &-30.000 \\
     7.0&  3.0 &  -5.110 &  -1.491 &  -1.127 &  -1.878 & -12.320 &-10.000 \\
     7.0& 30.0 &   1.702 &  -0.676 &   2.824 &   2.352 &   1.377 & -0.700 \\
    11.0&  1.0 &    -100 &    -100 &    -100 &    -100 &    -100 &-100\\
    11.0&  3.0 & -43.587 & -34.768 & -34.918 & -35.357 & -50.796 &-20.000 \\
    11.0& 30.0 &  -2.149 &  -3.775 &  -1.031 &  -1.497 &  -2.481 & -0.700 \\
\hline
\end{tabular}\\
\vspace{0.20cm}\\
\begin{tabular}{rr|rrrrrr}
\multicolumn{7}{c}{\normalsize{\hspace*{3.5cm}$^{  23}_{10}$ NE $_{ 13} \rightarrow \hspace*{0.2cm}^{  23}_{ 9}$ F  $_{ 14}$}}\\
\hline
Aden & T9& E-Cap &   E+Dec &  NuEn &  GamEn &  NeuEn &  PNEm \\
\hline
     3.0&  1.0 & -48.573 &    -100 & -49.145 & -50.452 & -88.849 &-40.000 \\
     3.0&  3.0 & -16.947 &    -100 & -17.015 & -18.220 & -30.581 &-10.000 \\
     3.0& 30.0 &   0.209 &    -100 &   1.228 &   0.510 &  -0.447 & -1.000 \\
     7.0&  1.0 & -42.753 &    -100 & -43.325 & -44.632 & -83.030 &-40.000 \\
     7.0&  3.0 & -15.232 &    -100 & -15.300 & -16.504 & -28.866 &-10.000 \\
     7.0& 30.0 &   0.213 &    -100 &   1.232 &   0.514 &  -0.444 & -1.000 \\
    11.0&  1.0 &   3.365 &    -100 &   4.299 &   3.925 &   2.203 & -1.000 \\
    11.0&  3.0 &   3.348 &    -100 &   4.281 &   3.916 &   2.242 & -1.000 \\
    11.0& 30.0 &   3.554 &    -100 &   4.715 &   4.014 &   3.204 & -0.700 \\
\hline
\end{tabular}\\
\vspace{0.20cm}\\
\noindent
\begin{tabular}{rr|rrrrrr}
\multicolumn{7}{c}{\normalsize{\hspace*{3.5cm}$^{  23}_{11}$ NA $_{ 12} \rightarrow \hspace*{0.2cm}^{  23}_{12}$ MG $_{ 11}$}}\\
\hline
Aden& T9&  E+Cap &   E-Dec &  ANuEn &  GamEn &  ProEn &  PPEm \\
\hline
     3.0&  1.0 & -25.398 & -30.285 & -25.346 & -32.274 & -66.169 &-40.000 \\
     3.0&  3.0 &  -9.267 & -14.614 &  -9.104 & -12.194 & -22.516 &-10.000 \\
     3.0& 30.0 &   0.843 &  -5.243 &   1.872 &   1.122 &   1.157 & -0.400 \\
     7.0&  1.0 & -31.218 & -32.059 & -31.154 & -38.094 & -71.989 &-40.000 \\
     7.0&  3.0 & -10.982 & -15.155 & -10.819 & -13.909 & -24.231 &-10.000 \\
     7.0& 30.0 &   0.839 &  -5.244 &   1.869 &   1.119 &   1.154 & -0.400 \\
    11.0&  1.0 &    -100 &    -100 &    -100 &    -100 &    -100 &-100\\
    11.0&  3.0 & -49.460 & -51.618 & -49.295 & -52.380 & -62.709 &-10.000 \\
    11.0& 30.0 &  -3.021 &  -8.558 &  -1.993 &  -2.742 &  -2.708 & -0.400 \\
\hline
\end{tabular}\\
\vspace{0.20cm}\\
\begin{tabular}{rr|rrrrrr}
\multicolumn{7}{c}{\normalsize{\hspace*{3.5cm}$^{  23}_{11}$ NA $_{ 12} \rightarrow \hspace*{0.2cm}^{  23}_{10}$ NE $_{ 13}$}}\\
\hline
Aden & T9& E-Cap &   E+Dec &  NuEn &  GamEn &  NeuEn &  PNEm \\
\hline
     3.0&  1.0 & -28.815 & -42.872 & -29.388 & -33.830 & -56.048 &-30.000 \\
     3.0&  3.0 & -11.199 & -16.138 & -11.269 & -12.271 & -19.785 & -8.000 \\
     3.0& 30.0 &   0.868 &  -4.570 &   1.914 &   0.795 &   1.062 & -0.300 \\
     7.0&  1.0 & -22.996 & -42.872 & -23.568 & -28.010 & -50.228 &-30.000 \\
     7.0&  3.0 &  -9.484 & -16.132 &  -9.554 & -10.555 & -18.070 & -8.000 \\
     7.0& 30.0 &   0.872 &  -4.568 &   1.918 &   0.798 &   1.066 & -0.300 \\
    11.0&  1.0 &   3.599 & -42.872 &   4.671 &   3.355 &   3.771 & -0.300 \\
    11.0&  3.0 &   3.644 & -16.132 &   4.734 &   3.456 &   3.736 & -0.400 \\
    11.0& 30.0 &   4.004 &  -4.375 &   5.240 &   3.814 &   4.426 & -0.200 \\
\hline
\end{tabular}\\
\vspace{0.20cm}\\
\noindent
\begin{tabular}{rr|rrrrrr}
\multicolumn{7}{c}{\normalsize{\hspace*{3.5cm}$^{  23}_{12}$ MG $_{ 11} \rightarrow \hspace*{0.2cm}^{  23}_{13}$ AL $_{ 10}$}}\\
\hline
Aden& T9&  E+Cap &   E-Dec &  ANuEn &  GamEn &  ProEn &  PPEm \\
\hline
     3.0&  1.0 & -66.187 &    -100 & -66.136 & -76.187 & -71.420 & -5.000 \\
     3.0&  3.0 & -22.538 &    -100 & -22.384 & -32.557 & -23.882 & -1.000 \\
     3.0& 30.0 &  -0.107 &    -100 &   0.892 & -11.004 &   0.467 & -0.060 \\
     7.0&  1.0 & -72.007 &    -100 & -71.956 & -82.007 & -77.240 & -5.000 \\
     7.0&  3.0 & -24.254 &    -100 & -24.100 & -34.273 & -25.597 & -1.000 \\
     7.0& 30.0 &  -0.110 &    -100 &   0.889 & -11.007 &   0.464 & -0.060 \\
    11.0&  1.0 &    -100 &    -100 &    -100 &    -100 &    -100 &-100\\
    11.0&  3.0 & -62.731 &    -100 & -62.577 & -72.750 & -64.075 & -1.000 \\
    11.0& 30.0 &  -3.973 &    -100 &  -2.974 & -14.869 &  -3.399 & -0.060 \\
\hline
\end{tabular}\\
\vspace{0.20cm}\\
\begin{tabular}{rr|rrrrrr}
\multicolumn{7}{c}{\normalsize{\hspace*{3.5cm}$^{  23}_{12}$ MG $_{ 11} \rightarrow \hspace*{0.2cm}^{  23}_{11}$ NA $_{ 12}$}}\\
\hline
Aden& T9&  E-Cap &   E+Dec &  NuEn &  GamEn &  ProEn &  PPEm \\
\hline
     3.0&  1.0 &  -5.093 &  -1.355 &  -1.135 &  -2.700 & -35.277 &-30.000 \\
     3.0&  3.0 &  -2.737 &  -1.429 &  -1.158 &  -2.596 & -12.286 &-10.000 \\
     3.0& 30.0 &   2.002 &  -1.312 &   3.094 &   2.407 &   2.240 & -0.400 \\
     7.0&  1.0 &  -1.269 &  -1.355 &  -0.524 &  -1.631 & -29.457 &-30.000 \\
     7.0&  3.0 &  -1.277 &  -1.426 &  -0.534 &  -1.504 & -10.572 &-10.000 \\
     7.0& 30.0 &   2.005 &  -1.311 &   3.097 &   2.410 &   2.244 & -0.400 \\
    11.0&  1.0 &   4.831 &  -1.355 &   6.058 &   5.310 &   4.822 & -0.500 \\
    11.0&  3.0 &   4.845 &  -1.426 &   6.064 &   5.343 &   4.860 & -0.400 \\
    11.0& 30.0 &   4.992 &  -1.178 &   6.281 &   5.335 &   5.427 & -0.200 \\
\hline
\end{tabular}\\
\vspace{0.20cm}\\
\noindent
\begin{tabular}{rr|rrrrrr}
\multicolumn{7}{c}{\normalsize{\hspace*{3.5cm}$^{  23}_{13}$ AL $_{ 10} \rightarrow \hspace*{0.2cm}^{  23}_{14}$ SI $_{  9}$}}\\
\hline
Aden& T9&  E+Cap &   E-Dec &  ANuEn &  GamEn &  ProEn &  PPEm \\
\hline
     3.0&  1.0 & -88.969 &    -100 & -88.918 & -91.686 &    -100 &-20.000 \\
     3.0&  3.0 & -30.018 &    -100 & -29.867 & -31.283 & -35.681 & -6.000 \\
     3.0& 30.0 &  -1.367 &    -100 &  -0.396 &  -2.357 &  -1.042 & -0.400 \\
     7.0&  1.0 & -94.788 &    -100 & -94.738 & -97.506 &    -100 &-20.000 \\
     7.0&  3.0 & -31.734 &    -100 & -31.583 & -32.998 & -37.396 & -6.000 \\
     7.0& 30.0 &  -1.370 &    -100 &  -0.399 &  -2.361 &  -1.045 & -0.400 \\
    11.0&  1.0 &    -100 &    -100 &    -100 &    -100 &    -100 &-100\\
    11.0&  3.0 & -70.211 &    -100 & -70.060 & -71.476 & -75.874 & -6.000 \\
    11.0& 30.0 &  -5.233 &    -100 &  -4.262 &  -6.224 &  -4.908 & -0.400 \\
\hline
\end{tabular}\\
\vspace{0.20cm}\\
\begin{tabular}{rr|rrrrrr}
\multicolumn{7}{c}{\normalsize{\hspace*{3.5cm}$^{  23}_{13}$ AL $_{ 10} \rightarrow \hspace*{0.2cm}^{  23}_{12}$ MG $_{ 11}$}}\\
\hline
Aden& T9&  E-Cap &   E+Dec &  NuEn &  GamEn &  ProEn &  PPEm \\
\hline
     3.0&  1.0 &  -4.419 &   0.550 &   1.265 &   0.589 &  -1.571 & -2.000 \\
     3.0&  3.0 &  -2.035 &   0.581 &   1.303 &   0.612 &  -1.506 & -2.000 \\
     3.0& 30.0 &   2.232 &   1.011 &   3.435 &   2.421 &   2.655 & -0.200 \\
     7.0&  1.0 &  -0.640 &   0.550 &   1.303 &   0.634 &  -0.943 & -1.000 \\
     7.0&  3.0 &  -0.597 &   0.582 &   1.342 &   0.657 &  -0.860 & -1.000 \\
     7.0& 30.0 &   2.236 &   1.011 &   3.439 &   2.425 &   2.658 & -0.200 \\
    11.0&  1.0 &   4.803 &   0.550 &   6.121 &   4.657 &   5.378 & -0.080 \\
    11.0&  3.0 &   4.809 &   0.582 &   6.128 &   4.674 &   5.384 & -0.080 \\
    11.0& 30.0 &   4.992 &   1.067 &   6.377 &   5.043 &   5.582 & -0.100 \\
\hline
\end{tabular}\\
\vspace{0.20cm}\\
\noindent
\begin{tabular}{rr|rrrrrr}
\multicolumn{7}{c}{\normalsize{\hspace*{3.5cm}$^{  24}_{10}$ NE $_{ 14} \rightarrow \hspace*{0.2cm}^{  24}_{11}$ NA $_{ 13}$}}\\
\hline
Aden & T9& E+Cap &   E-Dec &  ANuEn &  GamEn &  NeuEn &  PNEm \\
\hline
     3.0&  1.0 &  -6.100 &  -1.943 &  -1.908 &  -2.152 & -29.834 &-30.000 \\
     3.0&  3.0 &  -3.041 &  -1.951 &  -1.825 &  -2.106 & -10.673 & -8.000 \\
     3.0& 30.0 &   2.243 &   0.833 &   3.418 &   2.629 &   2.288 & -0.400 \\
     7.0&  1.0 & -11.920 &  -2.202 &  -2.282 &  -2.431 & -33.003 &-30.000 \\
     7.0&  3.0 &  -4.747 &  -2.124 &  -2.144 &  -2.342 & -12.208 &-10.000 \\
     7.0& 30.0 &   2.239 &   0.832 &   3.415 &   2.625 &   2.285 & -0.400 \\
    11.0&  1.0 &    -100 &    -100 &    -100 &    -100 &    -100 &-100\\
    11.0&  3.0 & -43.224 & -36.655 & -36.788 & -37.273 & -49.281 &-10.000 \\
    11.0& 30.0 &  -1.607 &  -2.013 &  -0.401 &  -1.166 &  -1.568 & -0.500 \\
\hline
\end{tabular}\\
\vspace{0.20cm}\\
\begin{tabular}{rr|rrrrrr}
\multicolumn{7}{c}{\normalsize{\hspace*{3.5cm}$^{  24}_{10}$ NE $_{ 14} \rightarrow \hspace*{0.2cm}^{  24}_{ 9}$ F  $_{ 15}$}}\\
\hline
Aden & T9& E-Cap &   E+Dec &  NuEn &  GamEn &  NeuEn &  PNEm \\
\hline
     3.0&  1.0 & -74.577 &    -100 & -75.155 & -76.496 & -94.079 &-20.000 \\
     3.0&  3.0 & -25.289 &    -100 & -25.373 & -26.108 & -32.064 & -6.000 \\
     3.0& 30.0 &   0.351 &    -100 &   1.353 &   0.149 &   0.333 & -0.400 \\
     7.0&  1.0 & -68.757 &    -100 & -69.335 & -70.677 & -88.260 &-20.000 \\
     7.0&  3.0 & -23.574 &    -100 & -23.657 & -24.392 & -30.348 & -6.000 \\
     7.0& 30.0 &   0.354 &    -100 &   1.357 &   0.153 &   0.337 & -0.400 \\
    11.0&  1.0 &   2.506 &    -100 &   3.300 &   2.665 &   1.020 & -1.000 \\
    11.0&  3.0 &   2.511 &    -100 &   3.308 &   2.669 &   1.036 & -1.000 \\
    11.0& 30.0 &   3.755 &    -100 &   4.905 &   3.515 &   3.911 & -0.300 \\
\hline
\end{tabular}\\
\vspace{0.20cm}\\
\noindent
\begin{tabular}{rr|rrrrrr}
\multicolumn{7}{c}{\normalsize{\hspace*{3.5cm}$^{  24}_{11}$ NA $_{ 13} \rightarrow \hspace*{0.2cm}^{  24}_{12}$ MG $_{ 12}$}}\\
\hline
Aden& T9&  E+Cap &   E-Dec &  ANuEn &  GamEn &  ProEn &  PPEm \\
\hline
     3.0&  1.0 &  -7.912 &  -4.094 &  -3.643 &  -4.259 & -37.748 &-30.000 \\
     3.0&  3.0 &  -4.595 &  -2.687 &  -2.178 &  -3.751 & -13.236 &-10.000 \\
     3.0& 30.0 &   1.434 &  -2.030 &   2.508 &   2.111 &   1.810 & -0.300 \\
     7.0&  1.0 & -13.732 &  -4.160 &  -3.689 &  -4.739 & -43.386 &-40.000 \\
     7.0&  3.0 &  -6.303 &  -2.704 &  -2.208 &  -4.561 & -14.951 &-10.000 \\
     7.0& 30.0 &   1.430 &  -2.031 &   2.504 &   2.108 &   1.807 & -0.300 \\
    11.0&  1.0 &    -100 & -95.948 & -96.548 &    -100 &    -100 &-60.000 \\
    11.0&  3.0 & -44.780 & -34.498 & -34.648 & -42.092 & -53.428 &-20.000 \\
    11.0& 30.0 &  -2.426 &  -5.226 &  -1.355 &  -1.745 &  -2.054 & -0.300 \\
\hline
\end{tabular}\\
\vspace{0.20cm}\\
\begin{tabular}{rr|rrrrrr}
\multicolumn{7}{c}{\normalsize{\hspace*{3.5cm}$^{  24}_{11}$ NA $_{ 13} \rightarrow \hspace*{0.2cm}^{  24}_{10}$ NE $_{ 14}$}}\\
\hline
Aden & T9& E-Cap &   E+Dec &  NuEn &  GamEn &  NeuEn &  PNEm \\
\hline
     3.0&  1.0 & -19.775 &    -100 & -20.335 & -24.401 & -65.056 &-40.000 \\
     3.0&  3.0 &  -8.388 &    -100 &  -8.436 &  -8.916 & -22.275 &-10.000 \\
     3.0& 30.0 &   1.204 &    -100 &   2.250 &   1.787 &   0.753 & -1.000 \\
     7.0&  1.0 & -13.955 &    -100 & -14.515 & -18.581 & -59.236 &-40.000 \\
     7.0&  3.0 &  -6.673 &    -100 &  -6.720 &  -7.200 & -20.560 &-10.000 \\
     7.0& 30.0 &   1.208 &    -100 &   2.254 &   1.791 &   0.756 & -1.000 \\
    11.0&  1.0 &   3.907 &    -100 &   5.093 &   4.143 &   2.979 & -1.000 \\
    11.0&  3.0 &   3.949 &    -100 &   5.132 &   4.222 &   3.146 & -1.000 \\
    11.0& 30.0 &   4.296 &    -100 &   5.551 &   4.843 &   4.289 & -0.700 \\
\hline
\end{tabular}\\
\vspace{0.20cm}\\
\noindent
\begin{tabular}{rr|rrrrrr}
\multicolumn{7}{c}{\normalsize{\hspace*{3.5cm}$^{  24}_{12}$ MG $_{ 12} \rightarrow \hspace*{0.2cm}^{  24}_{13}$ AL $_{ 11}$}}\\
\hline
Aden& T9&  E+Cap &   E-Dec &  ANuEn &  GamEn &  ProEn &  PPEm \\
\hline
     3.0&  1.0 & -75.324 &    -100 & -75.272 & -77.788 & -85.000 & -9.000 \\
     3.0&  3.0 & -25.644 &    -100 & -25.488 & -25.863 & -28.377 & -2.000 \\
     3.0& 30.0 &   0.118 &    -100 &   1.124 &  -0.587 &   0.630 & -0.100 \\
     7.0&  1.0 & -81.144 &    -100 & -81.092 & -83.608 & -90.820 & -9.000 \\
     7.0&  3.0 & -27.359 &    -100 & -27.204 & -27.578 & -30.093 & -2.000 \\
     7.0& 30.0 &   0.114 &    -100 &   1.121 &  -0.591 &   0.626 & -0.100 \\
    11.0&  1.0 &    -100 &    -100 &    -100 &    -100 &    -100 &-100\\
    11.0&  3.0 & -65.837 &    -100 & -65.681 & -66.056 & -68.570 & -2.000 \\
    11.0& 30.0 &  -3.748 &    -100 &  -2.742 &  -4.453 &  -3.236 & -0.100 \\
\hline
\end{tabular}\\
\vspace{0.20cm}\\
\begin{tabular}{rr|rrrrrr}
\multicolumn{7}{c}{\normalsize{\hspace*{3.5cm}$^{  24}_{12}$ MG $_{ 12} \rightarrow \hspace*{0.2cm}^{  24}_{11}$ NA $_{ 13}$}}\\
\hline
Aden & T9& E-Cap &   E+Dec &  NuEn &  GamEn &  NeuEn &  PNEm \\
\hline
     3.0&  1.0 & -36.771 & -43.563 & -37.140 & -38.008 & -69.758 &-30.000 \\
     3.0&  3.0 & -13.309 & -16.989 & -13.356 & -13.374 & -24.019 &-10.000 \\
     3.0& 30.0 &   1.253 &  -3.275 &   2.313 &   1.596 &   1.251 & -0.500 \\
     7.0&  1.0 & -31.160 & -43.563 & -31.697 & -32.188 & -63.944 &-30.000 \\
     7.0&  3.0 & -11.611 & -16.981 & -11.659 & -11.659 & -22.305 &-10.000 \\
     7.0& 30.0 &   1.257 &  -3.274 &   2.316 &   1.599 &   1.254 & -0.500 \\
    11.0&  1.0 &   4.122 & -43.563 &   5.209 &   4.457 &   3.134 & -1.000 \\
    11.0&  3.0 &   4.122 & -16.981 &   5.210 &   4.456 &   3.136 & -1.000 \\
    11.0& 30.0 &   4.346 &  -3.104 &   5.598 &   4.623 &   4.545 & -0.400 \\
\hline
\end{tabular}\\
\vspace{0.20cm}\\

\noindent
\begin{tabular}{rr|rrrrrr}
\multicolumn{7}{c}{\normalsize{\hspace*{3.5cm}$^{  24}_{13}$ AL $_{ 11} \rightarrow \hspace*{0.2cm}^{  24}_{14}$ SI $_{ 10}$}}\\
\hline
Aden& T9&  E+Cap &   E-Dec &  ANuEn &  GamEn &  ProEn &  PPEm \\
\hline
     3.0&  1.0 & -60.679 &    -100 & -60.628 & -67.475 & -82.382 &-20.000 \\
     3.0&  3.0 & -21.845 &    -100 & -21.691 & -24.042 & -28.842 & -7.000 \\
     3.0& 30.0 &  -0.859 &    -100 &   0.116 &  -2.854 &  -0.135 & -0.010 \\
     7.0&  1.0 & -66.499 &    -100 & -66.448 & -73.295 & -88.202 &-20.000 \\
     7.0&  3.0 & -23.560 &    -100 & -23.407 & -25.758 & -30.557 & -7.000 \\
     7.0& 30.0 &  -0.862 &    -100 &   0.112 &  -2.857 &  -0.139 & -0.010 \\
    11.0&  1.0 &    -100 &    -100 &    -100 &    -100 &    -100 &-100\\
    11.0&  3.0 & -62.038 &    -100 & -61.884 & -64.235 & -69.034 & -7.000 \\
    11.0& 30.0 &  -4.725 &    -100 &  -3.751 &  -6.720 &  -4.002 & -0.010 \\
\hline
\end{tabular}\\
\vspace{0.20cm}\\
\begin{tabular}{rr|rrrrrr}
\multicolumn{7}{c}{\normalsize{\hspace*{3.5cm}$^{  24}_{13}$ AL $_{ 11} \rightarrow \hspace*{0.2cm}^{  24}_{12}$ MG $_{ 12}$}}\\
\hline
Aden& T9&  E-Cap &   E+Dec &  NuEn &  GamEn &  ProEn &  PPEm \\
\hline
     3.0&  1.0 &  -4.540 &  -0.288 &   0.132 &   0.614 &  -5.735 & -5.000 \\
     3.0&  3.0 &  -2.158 &  -0.212 &   0.256 &   0.679 &  -3.587 & -3.000 \\
     3.0& 30.0 &   2.320 &   0.259 &   3.474 &   2.987 &   2.795 & -0.300 \\
     7.0&  1.0 &  -0.758 &  -0.288 &   0.372 &   0.747 &  -2.444 & -2.000 \\
     7.0&  3.0 &  -0.719 &  -0.211 &   0.467 &   0.800 &  -2.137 & -2.000 \\
     7.0& 30.0 &   2.324 &   0.260 &   3.477 &   2.990 &   2.799 & -0.300 \\
    11.0&  1.0 &   4.734 &  -0.288 &   6.030 &   5.328 &   5.242 & -0.300 \\
    11.0&  3.0 &   4.780 &  -0.211 &   6.071 &   5.318 &   5.322 & -0.200 \\
    11.0& 30.0 &   5.170 &   0.364 &   6.513 &   5.632 &   5.843 & -0.200 \\
\hline
\end{tabular}\\
\vspace{0.20cm}\\
\noindent
\begin{tabular}{rr|rrrrrr}
\multicolumn{7}{c}{\normalsize{\hspace*{3.5cm}$^{  24}_{14}$ SI $_{ 10} \rightarrow \hspace*{0.2cm}^{  24}_{15}$ P  $_{  9}$}}\\
\hline
Aden& T9&  E+Cap &   E-Dec &  ANuEn &  GamEn &  ProEn &  PPEm \\
\hline
     3.0&  1.0 &    -100 &    -100 &    -100 &    -100 &    -100 &-100\\
     3.0&  3.0 & -38.412 &    -100 & -38.262 &    -100 & -39.052 &  0.000 \\
     3.0& 30.0 &  -1.802 &    -100 &  -0.842 &    -100 &  -1.532 &  0.000 \\
     7.0&  1.0 &    -100 &    -100 &    -100 &    -100 &    -100 &-100\\
     7.0&  3.0 & -40.127 &    -100 & -39.977 &    -100 & -40.768 &  0.000 \\
     7.0& 30.0 &  -1.805 &    -100 &  -0.845 &    -100 &  -1.535 &  0.000 \\
    11.0&  1.0 &    -100 &    -100 &    -100 &    -100 &    -100 &-100\\
    11.0&  3.0 & -78.604 &    -100 & -78.455 &    -100 & -79.245 &  0.000 \\
    11.0& 30.0 &  -5.669 &    -100 &  -4.708 &    -100 &  -5.398 &  0.000 \\
\hline
\end{tabular}\\
\vspace{0.20cm}\\
\begin{tabular}{rr|rrrrrr}
\multicolumn{7}{c}{\normalsize{\hspace*{3.5cm}$^{  24}_{14}$ SI $_{ 10} \rightarrow \hspace*{0.2cm}^{  24}_{13}$ AL $_{ 11}$}}\\
\hline
Aden& T9&  E-Cap &   E+Dec &  NuEn &  GamEn &  ProEn &  PPEm \\
\hline
     3.0&  1.0 &  -4.034 &   0.792 &   1.408 &   0.681 &   0.467 & -0.400 \\
     3.0&  3.0 &  -1.675 &   0.792 &   1.410 &   0.681 &   0.472 & -0.400 \\
     3.0& 30.0 &   2.463 &   0.705 &   3.632 &   1.730 &   3.259 & -0.060 \\
     7.0&  1.0 &  -0.269 &   0.792 &   1.474 &   0.703 &   0.594 & -0.400 \\
     7.0&  3.0 &  -0.242 &   0.792 &   1.479 &   0.704 &   0.609 & -0.400 \\
     7.0& 30.0 &   2.467 &   0.705 &   3.636 &   1.733 &   3.263 & -0.060 \\
    11.0&  1.0 &   5.182 &   0.792 &   6.501 &   4.226 &   6.056 & -0.040 \\
    11.0&  3.0 &   5.183 &   0.792 &   6.502 &   4.226 &   6.057 & -0.040 \\
    11.0& 30.0 &   5.250 &   0.792 &   6.621 &   4.269 &   6.133 & -0.030 \\
\hline
\end{tabular}\\
\vspace{0.20cm}\\
\noindent
\begin{tabular}{rr|rrrrrr}
\multicolumn{7}{c}{\normalsize{\hspace*{3.5cm}$^{  25}_{10}$ NE $_{ 15} \rightarrow \hspace*{0.2cm}^{  25}_{11}$ NA $_{ 14}$}}\\
\hline
Aden & T9& E+Cap &   E-Dec &  ANuEn &  GamEn &  NeuEn &  PNEm \\
\hline
     3.0&  1.0 &  -4.958 &   0.209 &   0.590 &   0.673 & -20.067 &-20.000 \\
     3.0&  3.0 &  -1.912 &   0.219 &   0.591 &   0.727 &  -7.380 & -8.000 \\
     3.0& 30.0 &   2.444 &   0.843 &   3.628 &   3.080 &   1.961 & -1.000 \\
     7.0&  1.0 & -10.778 &   0.160 &   0.523 &   0.610 & -21.564 &-20.000 \\
     7.0&  3.0 &  -3.616 &   0.176 &   0.528 &   0.670 &  -9.090 &-10.000 \\
     7.0& 30.0 &   2.441 &   0.843 &   3.625 &   3.077 &   1.957 & -1.000 \\
    11.0&  1.0 &    -100 & -84.299 & -84.893 & -85.769 &    -100 &-50.000 \\
    11.0&  3.0 & -42.094 & -28.652 & -28.785 & -29.760 & -47.558 &-20.000 \\
    11.0& 30.0 &  -1.405 &  -2.004 &  -0.204 &  -0.761 &  -1.900 & -1.000 \\
\hline
\end{tabular}\\
\vspace{0.20cm}\\
\begin{tabular}{rr|rrrrrr}
\multicolumn{7}{c}{\normalsize{\hspace*{3.5cm}$^{  25}_{10}$ NE $_{ 15} \rightarrow \hspace*{0.2cm}^{  25}_{ 9}$ F  $_{ 16}$}}\\
\hline
Aden & T9& E-Cap &   E+Dec &  NuEn &  GamEn &  NeuEn &  PNEm \\
\hline
     3.0&  1.0 & -70.493 &    -100 & -71.074 & -72.018 & -98.663 &-30.000 \\
     3.0&  3.0 & -23.809 &    -100 & -23.901 & -24.993 & -33.547 &-10.000 \\
     3.0& 30.0 &  -0.096 &    -100 &   0.888 &  -0.103 &  -0.748 & -1.000 \\
     7.0&  1.0 & -64.673 &    -100 & -65.254 & -66.198 & -92.843 &-30.000 \\
     7.0&  3.0 & -22.093 &    -100 & -22.186 & -23.277 & -31.832 &-10.000 \\
     7.0& 30.0 &  -0.092 &    -100 &   0.892 &  -0.099 &  -0.745 & -1.000 \\
    11.0&  1.0 &   2.589 &    -100 &   3.307 &   3.032 &   1.955 & -0.900 \\
    11.0&  3.0 &   2.728 &    -100 &   3.506 &   3.090 &   1.993 & -1.000 \\
    11.0& 30.0 &   3.460 &    -100 &   4.554 &   3.587 &   3.033 & -0.800 \\
\hline
\end{tabular}\\
\vspace{0.20cm}\\
\noindent
\begin{tabular}{rr|rrrrrr}
\multicolumn{7}{c}{\normalsize{\hspace*{3.5cm}$^{  25}_{11}$ NA $_{ 14} \rightarrow \hspace*{0.2cm}^{  25}_{12}$ MG $_{ 13}$}}\\
\hline
Aden & T9& E+Cap &   E-Dec &  ANuEn &  GamEn &  NeuEn &  PNEm \\
\hline
     3.0&  1.0 &  -6.350 &  -1.700 &  -1.512 &  -1.653 & -23.850 &-20.000 \\
     3.0&  3.0 &  -3.399 &  -1.599 &  -1.314 &  -1.719 &  -9.305 & -7.000 \\
     3.0& 30.0 &   1.836 &  -0.381 &   2.948 &   2.134 &   2.333 & -0.200 \\
     7.0&  1.0 & -12.169 &  -1.831 &  -1.691 &  -1.824 & -29.670 &-30.000 \\
     7.0&  3.0 &  -5.104 &  -1.669 &  -1.415 &  -1.844 & -11.000 & -9.000 \\
     7.0& 30.0 &   1.833 &  -0.382 &   2.944 &   2.131 &   2.330 & -0.200 \\
    11.0&  1.0 &    -100 &    -100 &    -100 &    -100 &    -100 &-100\\
    11.0&  3.0 & -43.581 & -35.909 & -36.058 & -36.936 & -48.025 &-10.000 \\
    11.0& 30.0 &  -2.020 &  -3.529 &  -0.910 &  -1.696 &  -1.527 & -0.200 \\
\hline
\end{tabular}\\
\vspace{0.20cm}\\
\begin{tabular}{rr|rrrrrr}
\multicolumn{7}{c}{\normalsize{\hspace*{3.5cm}$^{  25}_{11}$ NA $_{ 14} \rightarrow \hspace*{0.2cm}^{  25}_{10}$ NE $_{ 15}$}}\\
\hline
Aden & T9& E-Cap &   E+Dec &  NuEn &  GamEn &  NeuEn &  PNEm \\
\hline
     3.0&  1.0 & -42.701 & -49.296 & -43.277 & -44.114 & -64.570 &-20.000 \\
     3.0&  3.0 & -15.145 & -20.350 & -15.222 & -16.271 & -22.172 & -6.000 \\
     3.0& 30.0 &   0.983 &  -7.923 &   2.025 &   1.039 &   0.984 & -0.300 \\
     7.0&  1.0 & -36.881 & -49.296 & -37.457 & -38.295 & -58.750 &-20.000 \\
     7.0&  3.0 & -13.429 & -20.312 & -13.507 & -14.556 & -20.457 & -6.000 \\
     7.0& 30.0 &   0.987 &  -7.921 &   2.029 &   1.043 &   0.987 & -0.300 \\
    11.0&  1.0 &   3.297 & -49.296 &   4.323 &   3.179 &   3.165 & -0.500 \\
    11.0&  3.0 &   3.279 & -20.312 &   4.314 &   3.078 &   3.149 & -0.500 \\
    11.0& 30.0 &   4.124 &  -7.671 &   5.360 &   4.066 &   4.383 & -0.200 \\
\hline
\end{tabular}\\
\vspace{0.20cm}\\
\noindent
\begin{tabular}{rr|rrrrrr}
\multicolumn{7}{c}{\normalsize{\hspace*{3.5cm}$^{  25}_{12}$ MG $_{ 13} \rightarrow \hspace*{0.2cm}^{  25}_{13}$ AL $_{ 12}$}}\\
\hline
Aden& T9&  E+Cap &   E-Dec &  ANuEn &  GamEn &  ProEn &  PPEm \\
\hline
     3.0&  1.0 & -26.305 & -29.994 & -26.253 & -31.862 & -40.244 &-10.000 \\
     3.0&  3.0 &  -9.325 & -13.016 &  -9.162 & -11.236 & -14.110 & -4.000 \\
     3.0& 30.0 &   0.878 &  -3.839 &   1.925 &   0.357 &   1.425 & -0.200 \\
     7.0&  1.0 & -32.125 & -31.373 & -31.728 & -37.680 & -46.064 &-10.000 \\
     7.0&  3.0 & -11.040 & -13.477 & -10.877 & -12.943 & -15.826 & -4.000 \\
     7.0& 30.0 &   0.874 &  -3.840 &   1.922 &   0.354 &   1.422 & -0.200 \\
    11.0&  1.0 &    -100 &    -100 &    -100 &    -100 &    -100 &-100\\
    11.0&  3.0 & -49.517 & -50.430 & -49.332 & -51.368 & -54.303 & -4.000 \\
    11.0& 30.0 &  -2.985 &  -7.240 &  -1.939 &  -3.501 &  -2.440 & -0.200 \\
\hline
\end{tabular}\\
\vspace{0.20cm}\\
\begin{tabular}{rr|rrrrrr}
\multicolumn{7}{c}{\normalsize{\hspace*{3.5cm}$^{  25}_{12}$ MG $_{ 13} \rightarrow \hspace*{0.2cm}^{  25}_{11}$ NA $_{ 14}$}}\\
\hline
Aden & T9& E-Cap &   E+Dec &  NuEn &  GamEn &  NeuEn &  PNEm \\
\hline
     3.0&  1.0 & -26.599 & -33.245 & -27.165 & -30.662 & -71.128 &-40.000 \\
     3.0&  3.0 & -10.596 & -15.909 & -10.653 & -11.277 & -24.338 &-10.000 \\
     3.0& 30.0 &   1.000 &  -5.621 &   2.030 &   1.545 &   0.851 & -0.600 \\
     7.0&  1.0 & -20.779 & -33.244 & -21.345 & -24.843 & -65.309 &-40.000 \\
     7.0&  3.0 &  -8.881 & -15.890 &  -8.938 &  -9.562 & -22.623 &-10.000 \\
     7.0& 30.0 &   1.004 &  -5.620 &   2.034 &   1.549 &   0.854 & -0.600 \\
    11.0&  1.0 &   3.887 & -33.244 &   4.903 &   4.492 &   3.369 & -0.600 \\
    11.0&  3.0 &   3.888 & -15.890 &   4.908 &   4.489 &   3.380 & -0.600 \\
    11.0& 30.0 &   4.233 &  -5.437 &   5.436 &   4.723 &   4.365 & -0.400 \\
\hline
\end{tabular}\\
\vspace{0.20cm}\\
\noindent
\begin{tabular}{rr|rrrrrr}
\multicolumn{7}{c}{\normalsize{\hspace*{3.5cm}$^{  25}_{13}$ AL $_{ 12} \rightarrow \hspace*{0.2cm}^{  25}_{14}$ SI $_{ 11}$}}\\
\hline
Aden& T9&  E+Cap &   E-Dec &  ANuEn &  GamEn &  ProEn &  PPEm \\
\hline
     3.0&  1.0 & -68.455 &    -100 & -68.404 & -73.692 & -95.013 &-30.000 \\
     3.0&  3.0 & -23.060 &    -100 & -22.908 & -25.002 & -31.813 & -9.000 \\
     3.0& 30.0 &  -0.482 &    -100 &   0.502 &  -0.619 &  -0.134 & -0.400 \\
     7.0&  1.0 & -74.275 &    -100 & -74.224 & -79.512 &    -100 &-30.000 \\
     7.0&  3.0 & -24.776 &    -100 & -24.623 & -26.717 & -33.528 & -9.000 \\
     7.0& 30.0 &  -0.485 &    -100 &   0.499 &  -0.622 &  -0.138 & -0.400 \\
    11.0&  1.0 &    -100 &    -100 &    -100 &    -100 &    -100 &-100\\
    11.0&  3.0 & -63.253 &    -100 & -63.100 & -65.195 & -72.006 & -9.000 \\
    11.0& 30.0 &  -4.348 &    -100 &  -3.364 &  -4.485 &  -4.001 & -0.400 \\
\hline
\end{tabular}\\
\vspace{0.20cm}\\
\begin{tabular}{rr|rrrrrr}
\multicolumn{7}{c}{\normalsize{\hspace*{3.5cm}$^{  25}_{13}$ AL $_{ 12} \rightarrow \hspace*{0.2cm}^{  25}_{12}$ MG $_{ 13}$}}\\
\hline
Aden & T9& E-Cap &   E+Dec &  NuEn &  GamEn &  NeuEn &  PNEm \\
\hline
     3.0&  1.0 &  -4.841 &  -1.019 &  -0.766 &  -2.861 & -24.078 &-20.000 \\
     3.0&  3.0 &  -2.457 &  -1.022 &  -0.730 &  -2.692 &  -9.309 & -8.000 \\
     3.0& 30.0 &   1.683 &  -1.279 &   2.808 &   1.569 &   1.717 & -0.400 \\
     7.0&  1.0 &  -1.048 &  -1.019 &  -0.229 &  -1.901 & -18.261 &-20.000 \\
     7.0&  3.0 &  -1.012 &  -1.020 &  -0.185 &  -1.834 &  -7.593 & -7.000 \\
     7.0& 30.0 &   1.687 &  -1.278 &   2.812 &   1.572 &   1.721 & -0.400 \\
    11.0&  1.0 &   4.492 &  -1.019 &   5.785 &   4.272 &   4.412 & -0.400 \\
    11.0&  3.0 &   4.493 &  -1.020 &   5.786 &   4.274 &   4.421 & -0.400 \\
    11.0& 30.0 &   4.579 &  -1.120 &   5.904 &   4.374 &   4.842 & -0.200 \\
\hline
\end{tabular}\\
\vspace{0.20cm}\\
\noindent
\begin{tabular}{rr|rrrrrr}
\multicolumn{7}{c}{\normalsize{\hspace*{3.5cm}$^{  25}_{14}$ SI $_{ 11} \rightarrow \hspace*{0.2cm}^{  25}_{15}$ P  $_{ 10}$}}\\
\hline
Aden& T9&  E+Cap &   E-Dec &  ANuEn &  GamEn &  ProEn &  PPEm \\
\hline
     3.0&  1.0 & -97.041 &    -100 & -96.991 &    -100 & -99.393 & -2.000 \\
     3.0&  3.0 & -32.936 &    -100 & -32.786 & -43.202 & -33.454 & -0.300 \\
     3.0& 30.0 &  -1.590 &    -100 &  -0.623 & -12.913 &  -1.136 & -0.020 \\
     7.0&  1.0 &    -100 &    -100 &    -100 &    -100 &    -100 &-100\\
     7.0&  3.0 & -34.651 &    -100 & -34.501 & -44.917 & -35.170 & -0.300 \\
     7.0& 30.0 &  -1.593 &    -100 &  -0.626 & -12.917 &  -1.140 & -0.020 \\
    11.0&  1.0 &    -100 &    -100 &    -100 &    -100 &    -100 &-100\\
    11.0&  3.0 & -73.129 &    -100 & -72.978 & -83.394 & -73.647 & -0.300 \\
    11.0& 30.0 &  -5.457 &    -100 &  -4.489 & -16.780 &  -5.003 & -0.020 \\
\hline
\end{tabular}\\
\vspace{0.20cm}\\
\begin{tabular}{rr|rrrrrr}
\multicolumn{7}{c}{\normalsize{\hspace*{3.5cm}$^{  25}_{14}$ SI $_{ 11} \rightarrow \hspace*{0.2cm}^{  25}_{13}$ AL $_{ 12}$}}\\
\hline
Aden& T9&  E-Cap &   E+Dec &  NuEn &  GamEn &  ProEn &  PPEm \\
\hline
     3.0&  1.0 &  -4.201 &   0.992 &   1.742 &   0.755 &   0.343 & -1.000 \\
     3.0&  3.0 &  -1.870 &   0.975 &   1.727 &   0.734 &   0.355 & -1.000 \\
     3.0& 30.0 &   2.250 &   0.764 &   3.411 &   1.275 &   3.228 & -0.060 \\
     7.0&  1.0 &  -0.443 &   0.992 &   1.768 &   0.769 &   0.481 & -1.000 \\
     7.0&  3.0 &  -0.440 &   0.975 &   1.754 &   0.747 &   0.497 & -1.000 \\
     7.0& 30.0 &   2.253 &   0.764 &   3.415 &   1.278 &   3.231 & -0.060 \\
    11.0&  1.0 &   5.049 &   0.992 &   6.338 &   3.999 &   6.076 & -0.040 \\
    11.0&  3.0 &   5.041 &   0.975 &   6.329 &   3.957 &   6.073 & -0.040 \\
    11.0& 30.0 &   5.098 &   0.825 &   6.443 &   3.732 &   6.159 & -0.020 \\
\hline
\end{tabular}\\
\vspace{0.20cm}\\
\noindent
\begin{tabular}{rr|rrrrrr}
\multicolumn{7}{c}{\normalsize{\hspace*{3.5cm}$^{  26}_{11}$ NA $_{ 15} \rightarrow \hspace*{0.2cm}^{  26}_{12}$ MG $_{ 14}$}}\\
\hline
Aden & T9& E+Cap &   E-Dec &  ANuEn &  GamEn &  NeuEn &  PNEm \\
\hline
     3.0&  1.0 &  -5.742 &  -0.302 &   0.245 &   0.135 & -14.320 &-10.000 \\
     3.0&  3.0 &  -2.704 &  -0.386 &   0.117 &   0.167 &  -6.076 & -5.000 \\
     3.0& 30.0 &   2.075 &   0.218 &   3.201 &   2.530 &   2.481 & -0.200 \\
     7.0&  1.0 & -11.562 &  -0.327 &   0.217 &   0.087 & -20.048 &-20.000 \\
     7.0&  3.0 &  -4.409 &  -0.413 &   0.084 &   0.121 &  -7.780 & -7.000 \\
     7.0& 30.0 &   2.072 &   0.217 &   3.198 &   2.526 &   2.478 & -0.200 \\
    11.0&  1.0 &    -100 & -84.823 & -85.421 & -84.566 &    -100 &-50.000 \\
    11.0&  3.0 & -42.886 & -30.156 & -30.296 & -29.898 & -45.538 &-10.000 \\
    11.0& 30.0 &  -1.780 &  -2.857 &  -0.651 &  -1.265 &  -1.380 & -0.200 \\
\hline
\end{tabular}%
\\
\vspace{0.20cm}\\


\begin{thebibliography}{99}
\bibitem{1} H. V. Klapdor, Prog. Part. Nucl. Phys. {\bf 10}, 131 (1983)
\bibitem{2} H. V. Klapdor, Prog. Part. Nucl. Phys. {\bf 17}, 419 (1986)
\bibitem{3} H. V. Klapdor-Kleingrothaus, AIP Conf. Proc. {\bf 238}, 870 (1991)
\bibitem{4} K. Grotz and H. V. Klapdor, \textit {The Weak Interaction in Nuclear, Particle and Astrophysics} (Adam Hilger, Bristol/Philadelphia/New York, 1990)
\bibitem{5} H. V. Klapdor, J. Metzinger, and T. Oda, ATOMIC DATA AND NUCLEAR DATA TABLES {\bf 31}, 81 (1984)
\bibitem{6} A. Staudt, E. Bender, K. Muto, and H. V. Klapdor, ATOMIC DATA AND NUCLEAR DATA TABLES {\bf 44}, 79 (1990)
\bibitem{7} A. Staudt, E. Bender, K. Muto, and H. V. Klapdor, Z. Phys. A {\bf 334}, 47 (1989)
\bibitem{8} M. Hirsch, A. Staudt, K. Muto, and H. V. Klapdor, ATOMIC DATA AND NUCLEAR DATA TABLES {\bf 53}, 165 (1993)
\bibitem{9} H. Homma, E. Bender, M. Hirsch, K. Muto, H. V. Klapdor-Kleingrothaus, and T. Oda, Phys. Rev. C {\bf 54}, 2972 (1996)
\bibitem{10} M. Takahara, M. Hino, T. Oda, K. Muto, A. A. Walters, P. W. M. Glaudemans, and K. Sato, Nucl.  Phys. A {\bf 504}, 167 (1989)
\bibitem{11} C. J. Copi, and D. N. Schramm, Comments. Nucl. Part. Phys. {\bf 22}, 1 (1996)
\bibitem{12} M. Politano, S. Starrfield, J. W. Truran, A. Weiss, and W. M. Sparks, Ap. J. {\bf 448}, 807 (1995)
\bibitem{13} P. Leleux, in Nuclei in the Cosmos IV, eds. J. G\"{o}rres, G. Mathews, S. Shore, and M. Wiescher, Nucl. Phys. (1997) in press.
\bibitem{14} F. K\"{a}ppeler, H. Beer, and K. Wisshak, Rep. Prog. Phys. {\bf 52}, 945 (1989)
\bibitem{15} B. S. Meyer, Ann. Rev. Astron. Astrophys. {\bf 32}, 153 (1994) 
\bibitem{16} N. Prantzos, M. Hashimoto, M. Rayet, and M. Arnould, Astron. Astrophys. {\bf 238}, 455 (1990)
\bibitem{17} M. Howard, B. Meyer, and S. E. Woosley, Ap. J. {\bf 373}, L5 (1991)
\bibitem{18} G. M. Fuller, W. A. Fowler, and M. J. Newman, Astrophys. J. Suppl. {\ bf 42}, 447 (1980); {\bf 48}, 279 (1982); Astrophys. J. {\bf 252}, 715 (1982)
\bibitem{29} B. A. Brown, W. Chung, and B. H. Wildenthal, Phys. Rev. Lett. {\bf 40}, 1631 (1978)
\bibitem{20} K. Takahashi, M. F. El Eid, and W. Hillebrandt, Astron. Astrophys. {\bf 67}, 185 (1978)
\bibitem{21} T. Oda, M. Hino, K. Muto, M. Takahara, and K. Sato, ATOMIC DATA AND NUCLEAR DATA TABLES {\bf 56}, 231 (1994)
\bibitem{22} B. H. Wildenthal, \textit {Progress in Particle and Nuclear Physics}, edited by D. H. Wilkinson (Pergamon, Oxford, 1984), Vol. 11, p.5
\bibitem{23} B. A. Brown, and B. H. Wildenthal, ATOMIC DATA AND NUCLEAR DATA TABLES {\bf 33}, 347 (1985)
\bibitem{24} E. Bender, K. Muto, and H. V. Klapdor, Phys. Lett. B {\bf 208}, 53 (1988)
\bibitem{25} K. Muto, E. Bender, T. Oda, and H. V. Klapdor, Z. Phys. A {\bf 341}, 407 (1992)
\bibitem{26} N. B.  Gove, and M. J. Martin, NUCLEAR DATA TABLES {\bf 10}, 205 (1971)
\bibitem{27} K. Muto, E. Bender, and H. V. Klapdor, Z. Phys. A {\bf 333}, 125 (1989)
\bibitem{28} K. Muto and H. V. Klapdor, in \textit{Neutrinos}, edited by H. V. Klapdor (Springer-Verlag, Berlin/Heidelberg/New York, 1988), pp. 183-237 
\bibitem{29} J. Randrup, Nucl. Phys. A {\bf 207}, 209 (1973); Licentiat thesis (unpublished)
\bibitem{30} K. Grotz, H. V. Klapdor, and J. Metzinger, Phys. Lett. B {\bf132}, 22 (1983)
\bibitem{31} H. V. Klapdor and K. Grotz, Phys. Lett. B {\bf142}, 323 (1984)
\bibitem{32} K. Grotz and H. V. Klapdor, \textit{The Weak Interaction in Nuclear, Particle and Astrophysics}, (Adam Hilger, Bristol/Philadelphia/New York, 1990)
\bibitem{33} D. R. Tilley, H. R. Weller, and C. M. Cheves, Nucl. Phys. A {\bf 565}, 1 (1993)
\bibitem{34} D. R. Tilley, H. R. Weller, C. M. Cheves, and R. M. Chasteler, Nucl. Phys. A {\bf 595}, 1 (1995)
\bibitem{35} F. Ajzenberg-Selove, Nucl. Phys. A {\bf 475}, 1 (1987)
\bibitem{36} P. M. Endt, Nucl. Phys. A {\bf 521}, 1 (1990)
\bibitem{37} G. Audi, and A. H. Wapstra, Nucl. Phys. A {\bf 595}, 409 (1995)
\bibitem{38} P. M\"oller and J. R. Nix, ATOMIC DATA AND NUCLEAR DATA TABLES {\bf 26}, 165 (1981)
\bibitem{39} M. Hirsch, A. Staudt, and H. V. Klapdor-Kleingrothaus, ATOMIC DATA AND NUCLEAR DATA TABLES {\bf 51}, 243 (1992)
\end{thebibliography}
\end{document}